\documentclass[a4paper,onecolumn,11pt,accepted=2021-09-06]{quantumarticle}
\pdfoutput=1
\usepackage[utf8]{inputenc}
\usepackage[english]{babel}
\usepackage[T1]{fontenc}
\usepackage{amsmath, amssymb, amsthm,}
\usepackage{hyperref}
\usepackage{bm}
\usepackage{tikz}
\usepackage{lipsum}

\usepackage[numbers,sort&compress]{natbib}

\def\tr{\mbox{tr}}

\def\1{\mathbbm{1}}

\def\eps{\varepsilon}
\DeclareUnicodeCharacter{00A0}{ }

\begin{document}

\title{Explicit asymptotic secret key rate of continuous-variable quantum key distribution with an arbitrary modulation}

\author{Aur\'elie Denys}
\affiliation{Inria, France}
%\orcid{}
\author{Peter Brown}
%\email{}
\orcid{0000-0001-9593-0136}
\affiliation{ENS Lyon, France}
\author{Anthony Leverrier}
\affiliation{Inria, France}
\orcid{0000-0002-6707-1458}
\email{anthony.leverrier@inria.fr}
\maketitle

\begin{abstract}
We establish an analytical lower bound on the asymptotic secret key rate of continuous-variable quantum key distribution with an arbitrary modulation of coherent states. Previously, such bounds were only available for protocols with a Gaussian modulation, and numerical bounds existed in the case of simple phase-shift-keying modulations. The latter bounds were obtained as a solution of convex optimization problems and our new analytical bound matches the results of Ghorai \textit{et al.} (2019), up to numerical precision. The more relevant case of quadrature amplitude modulation (QAM) could not be analyzed with the previous techniques, due to their large number of coherent states. Our bound shows that relatively small constellation sizes, with say 64 states, are essentially sufficient to obtain a performance close to a true Gaussian modulation and are therefore an attractive solution for large-scale deployment of continuous-variable quantum key distribution. We also derive similar bounds when the modulation consists of arbitrary states, not necessarily pure.
\end{abstract}

%%%%%%%%%%%%%%%%%%%%%%%%%%%%%%%%%%%%%%%%
\section{Introduction and main results}

Quantum key distribution (QKD) allows two distant parties with access to a quantum channel and an authenticated classical channel to share a secret key that can later encrypt classical messages \cite{SBC08,PAB20}. While the first protocols such as the celebrated Bennett-Brassard 84 protocol \cite{BB84} all relied on the exchange of discrete variables (DV) encoded for instance on the polarization of single photons, more recent protocols increasingly rely on a continuous-variable (CV) encoding in the quadratures of the quantified electromagnetic field, that benefits from state-of-the-art techniques in coherent optical telecommunication. 
This is particularly interesting since we are still at the early stages of a possible large-scale deployment of QKD, a deployment that would be greatly facilitated if the required technologies for QKD were fully compatible with standard telecom equipment. One can argue that CV QKD satisfies this description since the quantum part of the protocol consists in the exchange of coherent states modulated in phase-space and measurement with coherent detection. Roughly speaking, the main difference with classical coherent optical communication is that CV QKD works in the quantum regime with attenuated coherent states and low-noise detectors. 

CV QKD comes with some difficulties, however. In particular, security proofs for CV QKD are more complex since one cannot avoid a description in the full infinite-dimensional Fock space, while DV QKD protocols can more conveniently be described with Hilbert spaces of small dimension, making their theoretical analysis simpler. The crux of the problem is that one needs to be able to gather some statistics in the protocol (typically characterizing the level of correlations between the states sent by the first party, Alice, and the data obtained by the second party, Bob) and to infer how much information was obtained by a potential adversary controlling the quantum channel. In a DV protocol, the quantum channel acts on a low-dimensional quantum system and can therefore be relatively well constrained by measuring simple quantities like the quantum bit error rate. For a CV protocol on the other hand, the quantum channel acts on the full Fock space and is usually more difficult to characterize from easily accessible statistics.

At the moment, the only CV QKD protocols with a reasonably well-understood security proof are those where Alice prepares coherent states with a Gaussian modulation\footnote{Another CV QKD protocol with a full security proof relies on the exchange of squeezed states, combined with a homodyne measurement for Bob (that is, Bob measures only one of the two quadrature operators). This protocol is however significantly less practical than protocols with coherent states  \cite{CLV01,FFB12}.}. This means that for each use of the channel, she draws a random complex variable $\alpha$ from a Gaussian distribution and sends the coherent state $|\alpha\rangle = e^{-|\alpha|^2/2} \sum_{n=0}^\infty \frac{\alpha^n}{\sqrt{n!}}|n\rangle$ to Bob. If Bob's measurement is a heterodyne detection, this corresponds to the no-switching protocol \cite{WLB04}. The phase-space symmetries of this protocol allow one to apply  the Gaussian de Finetti theorem which asserts that Gaussian attacks are asymptotically optimal \cite{lev17,lev18}. In other words, forgetting for the moment about finite-size effects, one can simply assume that the unknown channel between Alice and Bob is the Gaussian channel compatible with the statistics observed by Alice and Bob.

Unfortunately, a Gaussian modulation is merely a theoretical idealization since in practice modulators have a finite range and precision, meaning that the true number of states possibly available is finite. For instance, if the modulator has 8 bits of precision, we get $2^8= 256$ values per quadrature and $2^{16} = 65\, 536$ possible coherent states. While this number certainly looks large, is it really the case that a CV QKD protocol with this many states automatically inherits the security guarantees derived for a Gaussian modulation? Ref.~\cite{KGW19} looked at this specific question and found that, modulo some mild additional assumptions, it seems likely that the asymptotic secret key rate would be close to that of the Gaussian modulation for constellations of size greater than 5000. The approach there is to show that if the constellation is sufficiently close to the Gaussian one, then it is possible to exploit continuity bounds on the secret key rate together with the established security proofs for the Gaussian modulation in order to get reasonable numerical bounds for the secret key rate, when the constellation is large enough. This method, however, does not seem well-suited to address the case of significantly smaller constellation sizes.

At the other end of the spectrum, it is tempting to drastically reduce the number of coherent states in the constellation to simplify as much as possible the hardware requirements of the protocols as well as the reconciliation procedure (where Alice and Bob extract a common raw key from their correlated data). Protocols with 2, 3 or 4 coherent states have been considered in the literature and are part of the general class of $M$-PSK (phase-shift keying) protocols where Alice sends coherent states of the form $|\alpha_k\rangle = |\alpha e^{2\pi i k/M}\rangle$ for some $\alpha >0$ \cite{HYA03, LKL04,HL07,LG09,ZHR09,SL10,BW18,MMS21,PP21}.
While $M=2$ or $3$ appear to be too small to yield good performance, the 4-PSK (also known as quadrature phase-shift keying, QPSK) modulation scheme has attracted some interest since it performs reasonably well, although quite far from a Gaussian modulation.
Until recently, before the works of Refs \cite{GGD19,LUL19}, all the security proofs for the QPSK protocol were restricted to the class of Gaussian attacks (meaning that the quantum channel is assumed to be Gaussian\footnote{In fact, the proofs only assumed that the quantum channel acted linearly on the annihilation and creation operators, possibly adding non-Gaussian noise.}); it is believed that such attacks are not optimal for these protocols. The strategy in both Refs \cite{GGD19,LUL19} consists in expressing the asymptotic secret key rate as a convex optimization problem, and more precisely a semidefinite program (SDP). The main difference between the two papers is that Ref.~\cite{GGD19} considers a linear objective function, while Ref.~\cite{LUL19} relies on a tighter nonlinear objective function. While the latter case is expected to give a better bound (at the price of being much more computationally intensive), the results cannot be directly compared since the models and assumptions for the error correction part of the protocol are very different (see Section \ref{sec:beta} for a discussion of this point). In both cases, a truncated version of the relevant SDP is solved numerically: this means that the operators are described in a truncated Fock space, spanned by Fock states with less than $N_{\max}$ photons, typically between 10 and 20 photons. Very recently, Ref.~\cite{UHJ21} showed how to get rid of this truncation by introducing extra constraints in the SDP, namely constraints on the fourth moments of the data obtained by Alice and Bob. If the approaches of \cite{GGD19,LUL19,UHJ21} can in principle be adapted to arbitrary modulation schemes, they are numerically intensive\footnote{For instance, the size of the matrices involved in the SDP in \cite{GGD19} scales like $MN_{\max}$, where $M$ is the number of states in the constellation and $N_{\max}$ is the dimension of the truncated Fock space. Going beyond $M=10$ seems very challenging. The approach of \cite{LUL19,UHJ21} is even more expensive since the objective functional is not linear.} and it is unlikely that they can indeed be easily applied beyond moderately small PSK modulations. In fact, Ref.~\cite{PP21} which only looks at the simpler case of Gaussian (hence likely non optimal) attacks comments that several hours of CPU time are needed to get an accurate bound on the secret key rate.

\paragraph{Results and open questions.}
A pressing open question in the field is therefore to obtain reasonably tight bounds for the asymptotic secret key rate of CV QKD with arbitrary modulation schemes, that can be easily computed, without relying on intensive computational methods. Without this, it seems rather hopeless to try to address the next important challenge which will concern the non-asymptotic regime. 
We solve this problem here: we give an explicit analytical formula for the asymptotic secret key rate of any CV QKD protocol. While we focus more on the case of heterodyne detection, our bounds work just as well for protocols with homodyne detection \cite{GG02}. Our formula matches the numerical bound from Ref.~\cite{GGD19} in the case of $M$-PSK modulation of coherent states (except in the regime of very low loss combined with high noise, which is not relevant for experiments) and recovers the known values in the case of a Gaussian modulation. 
Our results show that relatively small constellations of size 64, say, are essentially enough to get a performance close to the Gaussian modulation scheme. A major advantage of the quadrature amplitude modulation such as 64-QAM over QPSK (in addition to the much better secret key rate) is that it allows for implementations with large modulation variance, and therefore bypasses the need to work with an extremely low signal-to-noise ratio (SNR). 

Another advantage of our method is that our analytical formula allows one to address the issue of imperfect state preparation. More precisely, in a given protocol, Alice will never be able to prepare the exact states from the theoretical constellation, and will inevitably make some preparation errors. Quantifying their impact on the security is not trivial if one only has access to numerical bounds, but this becomes possible with analytical bounds by analyzing their dependence on the constellation. We show in Section \ref{sec:thermal} how to modify our bound if Alice sends some (potentially mixed) state $\tau_k$ instead of $|\alpha_k\rangle$. The same bounds also apply to the case of a modulation of single-mode squeezed states, although such protocols are less appealing from a practical point of view. 

Yet another advantage of easily computable bounds is that they will allow for a better optimization of the constellation. While the PSK modulation does not offer much freedom since the only parameters are the number of states and the amplitude $\alpha$ of the coherent states, more complex constellations can have many adjustable parameters: the coherent states can lie on a grid, but not necessarily, and one can also freely choose the probabilities associated to each state. In this paper, we focus on simple QAM with equidistant coherent states, and only compare two possible choices for the probability distribution (discrete Gaussian \textit{vs} binomial). While the precise form of the constellation does not seem to impact the performance too much for a 64-QAM or larger constellations, we expect that smaller constellations will need to be more carefully designed in order to optimize the secret key rate. Such optimizations should include considerations about error correction\footnote{A possibility would be to use a 32-QAM, but the reconciliation may be more complex since Alice does not choose the values of $\mathrm{Re}(\alpha)$ and $\mathrm{Im}(\alpha)$ independently in that case.}, and are also beyond the scope of this paper. 

A natural open question concerns the case of the QPSK modulation. For this specific choice of constellation, our results (which coincide with Ref.~\cite{GGD19}) appear much more pessimistic than those of Ref.~\cite{LUL19}. This is due in part to the different choice of objective function and it would be very interesting to understand whether an analytical bound much tighter than ours could be derived explicitly. For larger constellations, our bound is necessarily almost tight since it is very close to the (tight) bound corresponding to a Gaussian modulation (see Section \ref{sec:num}).

While we focus on one-way QKD protocols here for simplicity, we note that similar questions are relevant for measurement-device-independent protocols \cite{POS15}. In that case, both Alice and Bob are expected to send states with a possibly very fine, but discrete, constellation approaching a Gaussian modulation. It would be interesting to understand how to extend our results to this scenario.

The asymptotic secret key rate is an interesting figure of merit that is useful to easily compare various protocols, either DV or CV, under some given experimental conditions. However, it is not quite sufficient to assess the security of a given protocol. What is needed is in fact a composable security proof valid against general attacks, in the finite-size regime. Obtaining such a security proof has turned out to be quite challenging in the case of the Gaussian modulation with a proof based on a Gaussian de Finetti theorem \cite{lev17} while the asymptotic secret key rate formula was established more than 10 years earlier \cite{GC06,NGA06}. Similarly, we do not give a full composable security proof here, but show that probably the two most impacting finite-size effects (see discussion in Section \ref{sec:PE}), namely the parameter estimation procedure and the error reconciliation procedure (see discussion in Section \ref{sec:beta}), should not be significantly more difficult to handle than they are in the case of Gaussian modulation. \\

\paragraph{Structure of the paper.}
We describe the general form of CV QKD protocols with coherent states in Section \ref{sec:protocol}. 
We explain in Section \ref{sec:EB} how to compute the asymptotic secret key rate given by the Devetak-Winter bound thanks to an equivalent entanglement-based version of the protocol. In Section \ref{sec:def-sdp}, we define our main lower bound on the Devetak-Winter bound as the solution of a semidefinite program. We study this SDP in Section \ref{sec:primal} and establish an analytical lower bound on its value. This bound is our main technical contribution. In Sections \ref{sec:Gaussian} and \ref{sec:PSK}, we show how to recover the known bound for protocols with a Gaussian modulation and the known numerical bound for protocols with an $M$-PSK modulation. We discuss in Section \ref{sec:constellation} the choice of more complex modulation schemes, namely QAM. We show in Section \ref{sec:thermal} how to generalize our bound for protocols where Alice sends arbitrary states instead of coherent states. We address some important finite-size effects in Section \ref{sec:finite}, notably parameter estimation and the reconciliation procedure. Finally, we discuss some numerical results in Section \ref{sec:num}.

%%%%%%%%%%%%%%%%%%%%%%%%%%%%%%%%%%%%%%%%
\section{CV QKD protocols with an arbitrary modulation of coherent states}
\label{sec:protocol}

\paragraph{Modulation schemes.}
We consider the following Prepare-and-Measure (PM) protocol where Alice sends coherent states chosen from a discrete modulation to Bob, who measures them with coherent (heterodyne) detection\footnote{We could similarly focus on protocols with homodyne detection, but the advantage of heterodyne detection is that it is more symmetric in phase-space and security against general attacks might therefore be easier to analyze in that case.}. A heterodyne detection refers here to a double-homodyne detection, where Bob splits the signal on a balanced beamsplitter and measures the $\hat{x}$ quadrature of the first output mode and the $\hat{p}$ quadrature of the second output mode. 
The modulation scheme is defined by a set of coherent states $\{ |\alpha_k\rangle\}$, called the constellation, where a state $|\alpha_k\rangle$ is chosen with probability $p_k$. This information can be summarized by a density matrix $\tau$ given by the weighted mixture of coherent states, and corresponding to the average state sent by Alice:
\begin{align}\label{eqn:tau}
\tau := \sum_k p_k |\alpha_k\rangle \langle \alpha_k|.
\end{align}
Note that for any finite constellation, this state faithfully describes the modulation scheme since the coherent states $|\alpha_k\rangle$ are linearly independent (this will no longer be the case in general if Alice sends mixed states, \textit{e.g.} thermal states). An important parameter is the variance of the modulation. In this paper, we define the quadrature operators by $\hat{x} := \hat{a} + \hat{a}^\dag$ and $\hat{p} := -i (\hat{a}- \hat{a}^\dag)$, where $\hat{a}$ and $\hat{a}^\dag$ (resp.~$\hat{b}, \hat{b}^\dag$) are the annihilation and creation operators\footnote{When the context is clear, we will sometimes omit the hat on the operators and simply write $a, a^\dag$ instead of $\hat{a}, \hat{a}^\dag$.} on Alice's system (resp.~Bob's system), and get the commutation relation $[\hat{x},\hat{p}]=2i$. 
The covariance matrix $\Gamma_\tau$ of the state $\tau$ is defined by 
\begin{align*}
\Gamma_\tau  := \begin{bmatrix}
\langle \hat{x}^2\rangle_\tau & \frac{1}{2} \langle \{\hat{x}, \hat{p} \}\rangle_\tau \\
\frac{1}{2} \langle \{\hat{p}, \hat{x} \}\rangle_\tau & \langle  \hat{p}^2 \rangle_\tau 
 \end{bmatrix}
\end{align*}
where we assumed without loss of generality that the first moment of the displacement operator vanishes (this can always be enforced by a suitable translation in phase-space).
We have for instance $\frac{1}{2} (\langle \hat{x}^2\rangle_\tau +\langle \hat{p}^2\rangle_\tau)= \tr( \tau (1 + 2 \hat{a}^\dag \hat{a} + \hat{a}^2 + \hat{a}^{\dagger 2})) = 1 +2 \langle n \rangle$, where the average photon number $\langle n \rangle$ in the modulation is defined as
\[\langle n \rangle := \sum_k p_k |\alpha_k|^2.\]
It is also customary to refer to $2\langle n \rangle$ as the modulation variance $V_A$ so that $\frac{1}{2} (\langle \hat{x}^2\rangle_\tau +\langle \hat{p}^2\rangle_\tau)  = V_A +1$.

There are two main modulation schemes usually discussed in the literature: the Gaussian modulation and the $M$-PSK modulation. 
In the case of a Gaussian modulation of variance $1 +2\langle n \rangle$, the value of $\alpha$ is an arbitrary complex number chosen according to a Gaussian probability distribution, and the associated density matrix $\tau_{\mathrm{G}}$ is a thermal state:
\[ \tau_{\mathrm{G}} = \frac{1}{\pi \langle n \rangle} \int_{\mathbbm{C}} \exp\left( -\frac{1}{\langle n \rangle} |\alpha|^2\right) |\alpha\rangle \langle \alpha| d\alpha = \frac{1}{1+\langle n \rangle} \sum_{m=0}^\infty \left(\frac{\langle n \rangle}{1+\langle n \rangle}\right)^m |m\rangle \langle m |,\] 
where $|n\rangle := \frac{\hat{a}^{\dag n}}{\sqrt{n!}}|0\rangle$ is the Fock state with $n$ photons. 
In the $M$-PSK modulation case, Alice chooses uniformly at random a coherent state from the set $\{ |\alpha e^{2\pi i k/M}\rangle\}_{0 \leq k \leq M-1}$ where the modulation variance corresponds to $V_A = 2\alpha^2$. The corresponding mixture is 
\[ \tau_{{M\text{-PSK}}} = \frac{1}{M} \sum_{k=0}^{M-1} |\alpha e^{2\pi ik/M}\rangle \langle \alpha e^{2\pi ik/M}|.\]
Note that the case $M=4$, also referred to as quadrature phase-shift keying (QPSK), has been widely studied in the context of CV QKD. 
The Gaussian and $M$-PSK modulation schemes are discussed in more details in Sections \ref{sec:Gaussian} and \ref{sec:PSK}, respectively. 
\begin{figure}[!h]
\begin{center}
\includegraphics[trim=20 10 0 2,clip,scale=0.48]{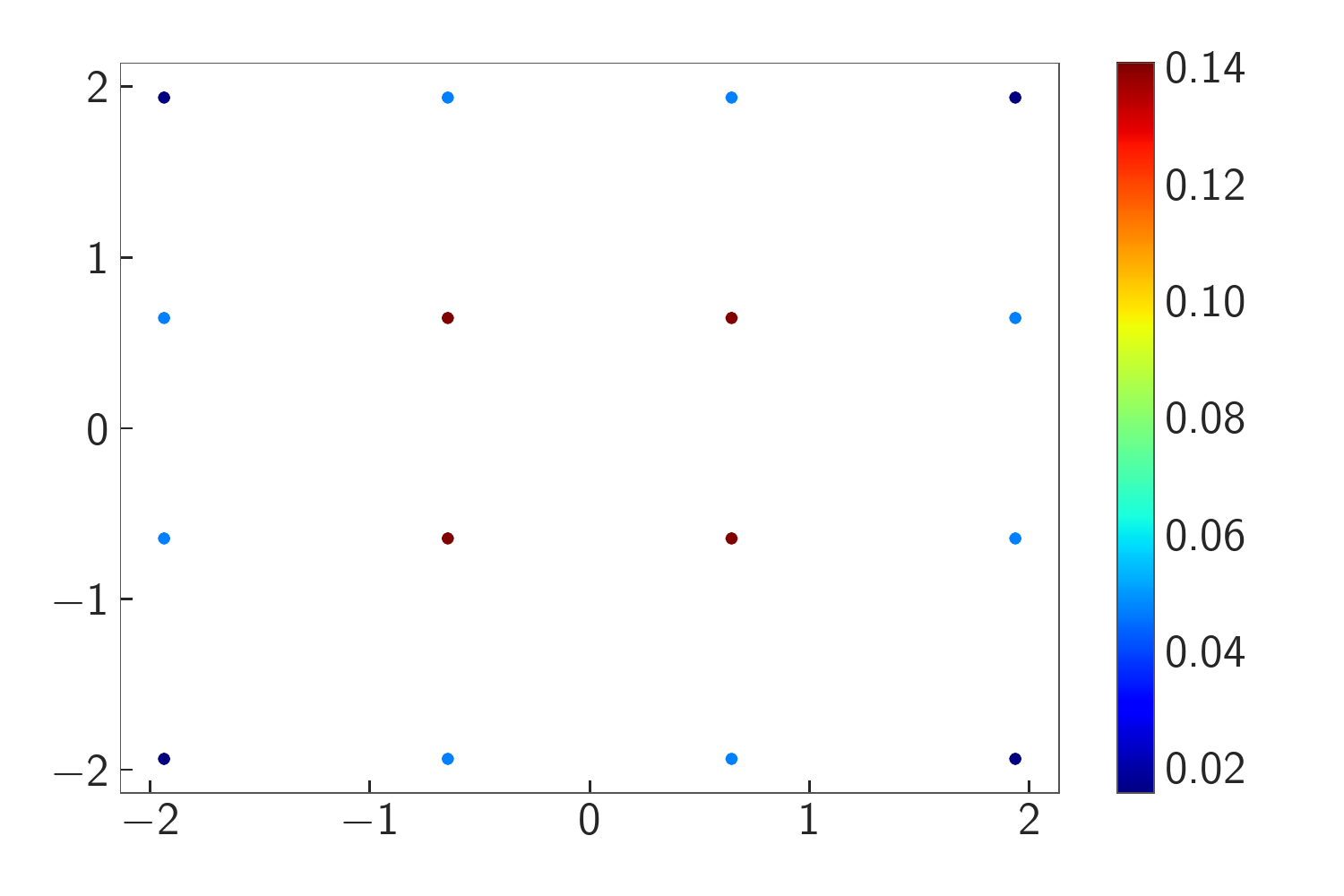}\includegraphics[trim=0 10 0 2,clip,scale=0.48]{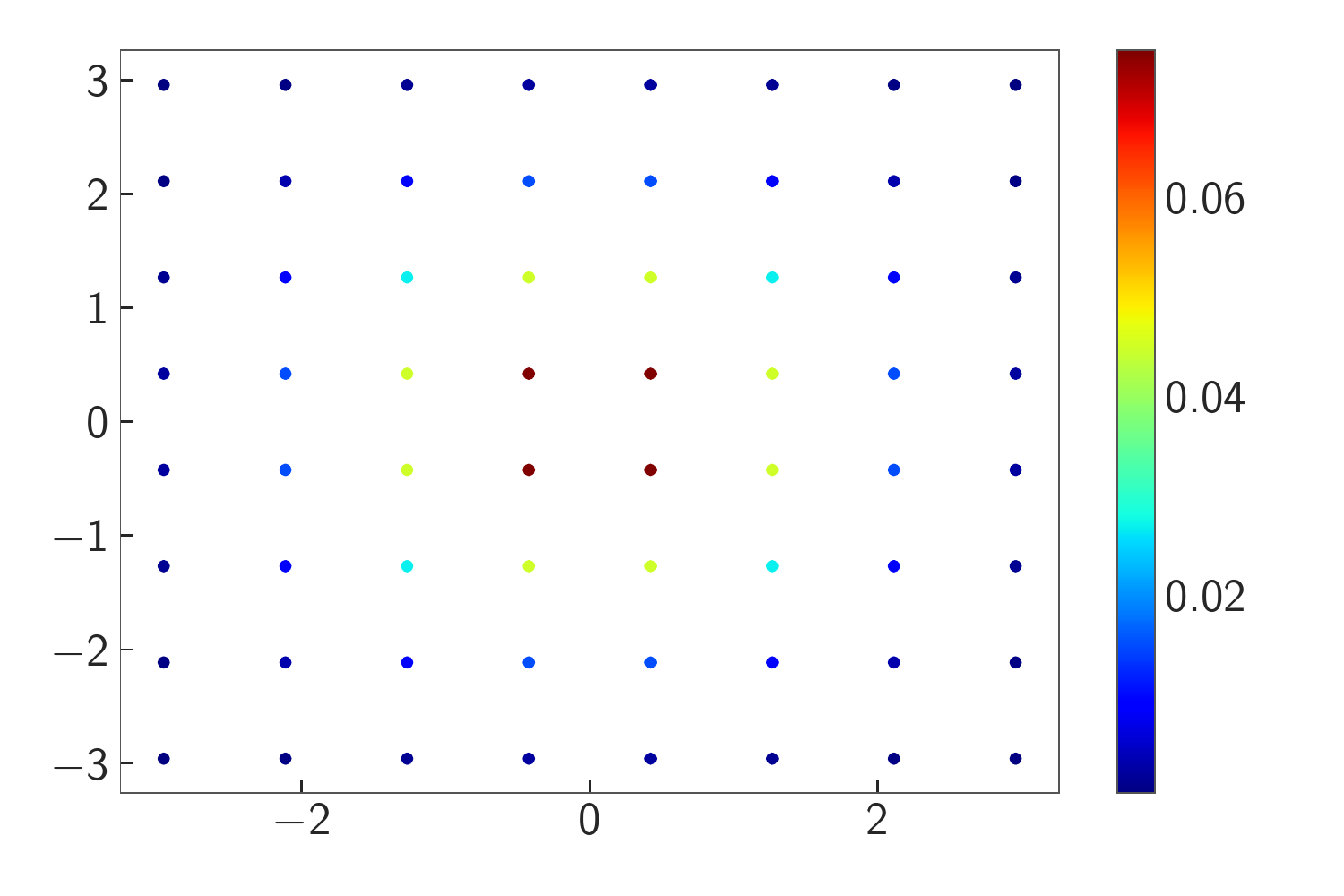}\end{center}
\caption{Constellations corresponding to a 16-QAM and a 64-QAM. Colors indicate the probabilities corresponding to each coherent state, following here a binomial distribution with $V_A=5$ (see Section \ref{sec:constellation} for details).}
\label{fig:QAM}
\end{figure}

In coherent optical communications, it is known that increasing the value of $M$ beyond 10, say, is not beneficial and that it is more efficient to switch instead to a different modulation scheme altogether. One such example is quadrature amplitude modulation (QAM) where the constellation typically consists of $M$ points distributed over a square grid (see Figure \ref{fig:QAM}). It is typical to consider $M$ to be a power of $4$, and we will indeed consider 4-QAM (which corresponds to QPSK), 16-QAM, 64-QAM, 256-QAM and 1024-QAM in this paper. 
Given that our proof technique will work better when a modulation scheme is closer to the Gaussian modulation, it is crucial that the $M$ points of the QAM are not chosen with a uniform probability distribution. Rather, we will consider probabilistic constellation shaping \cite{GJR17,JEM18} where each coordinate of the coherent state $|\alpha_k\rangle$ is chosen independently according to either a binomial or a Gaussian distribution (see Section \ref{sec:constellation} for details). More complex constellations are also possible.

\paragraph{The Prepare-and-Measure (PM) CV QKD protocol.}
Any QKD protocol consists of two main parts: a quantum part where Alice and Bob exchange quantum states and obtain correlated variables, and a classical post-processing procedure aiming at extracting two identical secret keys out of the correlated data. 
We have already described the first part. Alice and Bob repeat a large number of times the following: Alice chooses an index $k$ with probability $p_k$ and sends the corresponding coherent state $|\alpha_k\rangle$ to Bob through an untrusted quantum channel; Bob measures each incoming state with heterodyne detection\footnote{In a protocol with homodyne detection, Bob would only measure a random quadrature and afterwards inform Alice of his choice.} obtaining a complex number $\beta$. At the end of this first phase, Alice and Bob both hold a string of complex numbers. 
The goal of the second phase of the protocol is to use classical post-processing to transform these two strings into identical secret keys. It requires four steps: $(i)$ Bob discretizes his variables by choosing an appropriate binning of the complex plane\footnote{The bins should be small enough to guarantee that the reconciliation efficiency is close to 1.}; $(ii)$ in the reconciliation step, he sends some side-information to Alice \textit{via} the classical authenticated channel in order to help her guess Bob's string\footnote{We consider here the case, known as reverse reconciliation \cite{GG02b}, where the raw key corresponds to Bob's string since it always outperforms protocols where Alice's string is used as a raw key.}, (exploiting the side information together with her knowledge of the states she has sent); $(iii)$ Alice and Bob perform parameter estimation in order to bound how much information was possibly obtained by a malicious eavesdropper; and $(iv)$ they perform privacy amplification in order to obtain a shorter shared bit string completely unknown to the adversary.
All these steps must be carefully analyzed for a full security proof, but since our goal is the asymptotic regime, we will only mainly comment the reconciliation procedure and the parameter estimation step in Section \ref{sec:finite}.

%%%%%%%%%%%%%%%%%%%%%%%%%%%%%%%%%%%%%%%%
\section{Entanglement-Based protocol and Devetak-Winter bound}
\label{sec:EB}

In order to analyze the security of a PM protocol as defined in the previous section, the standard technique consists in defining an equivalent entanglement-based  (EB) version of the protocol, which only differs from the practical protocol in Alice's lab. Since both protocols are indistinguishable from the perspective of Bob and the adversary, they share the same security.

The EB version of the protocol is as follows: Alice prepares a bipartite state $|\Phi\rangle_{AA'}$, which is a purification of $\tau$, and measures the first mode in a basis that projects the second mode $A'$ onto the coherent states corresponding to the modulation scheme of the PM protocol. 
In this version, the second mode $A'$ is sent through the quantum channel $\mathcal{N}_{A' \to B}$ (controlled by the adversary), and Bob obtains the output mode $B$. We denote by $\rho_{AB} = (\mathrm{id}_A \otimes \mathcal{N}_{A' \to B}) (|\Phi\rangle\langle\Phi|_{AA'})$ the state shared by Alice and Bob after each use of the channel, where $\mathrm{id}_A$ stands for the identity channel acting on system $A$. In the present paper, we study so-called collective attacks in the asymptotic regime, and therefore assume that the channel is always the same (but unknown) during the protocol, which means that Alice and Bob share a large number of copies of the state $\rho_{AB}$. We note that collective attacks are usually optimal among all possible attacks in the asymptotic limit \cite{ren07}, and it therefore makes sense to consider these attacks here. 

The well-known Devetak-Winter bound gives the achievable secret key rate $K$ (per channel use) in this setup \cite{DW05}:
\begin{align}\label{eqn:DW}
K =  I(X;Y)- \sup_{\mathcal{N}: A'\to B} \chi(Y;E),
\end{align}
where $I(X;Y)$ is the mutual information between Alice and Bob's classical variables $X$ and $Y$ (which are complex variables in a protocol with heterodyne measurement, and real variables for homodyne measurement) and $\chi(Y;E)$ is the Holevo information between $Y$ and the quantum register $E$ of the adversary, with the supremum computed over all choices of channels $\mathcal{N}: A' \to B$ compatible with the statistics obtained by Alice and Bob during the parameter estimation phase of the PM protocol. The register $E$ of the adversary is introduced \textit{via} the isometric representation of the quantum channel, $U_{A' \to BE}$, which allows one to write a purification $\rho_{ABE}$ of $\rho_{AB}$:
\[ \rho_{ABE} = (\mathrm{id}_A \otimes \mathcal{U}_{A'\to BE})(|\Phi\rangle\langle\Phi|_{A A'}),\]
and $\rho_{AYE} = \mathcal{M}_{B \to Y}(\rho_{ABE})$ where the map $\mathcal{M}: B \to Y$ describes the (trusted) Gaussian measurement performed by Bob. In the case of a heterodyne measurement, it is given by
\[ \mathcal{M}(\rho_B) = \frac{1}{\pi} \int_{\mathbbm{C}} \langle \beta|\rho_B|\beta\rangle |\beta^{\mathrm{cl}}\rangle \langle \beta^{\mathrm{cl}}|_Y d\beta,\]
where $\{ |\beta^{\mathrm{cl}}\rangle\}$ is an infinite orthonormal family of states storing the value of the measurement outcome. The Holevo information $\chi(Y;E)$ is computed for the state $\rho_{AYE}$, and the supremum can also be computed over such states that are compatible with the statistics obtained in the parameter estimation step. 

In the finite-size regime, it is not quite possible for Alice and Bob to perfectly extract all their mutual information, and it is customary to replace $I(X;Y)$ by $\beta I(X;Y)$ where the reconciliation efficiency $\beta$ is a parameter that quantifies how much extra information Bob needs to send to Alice through the authenticated classical channel for her to correctly infer the value of $Y$. Modern techniques usually allow one to get $\beta \geq 0.95$. In any case, the value of $\beta I(X;Y)$ can be observed during a given protocol
Bounding the value of $\sup_{\mathcal{N}: A'\to B} \chi(Y;E)$ is more complicated, however, since it involves an optimization over a family of infinite-dimensional quantum channels. A very useful tool in this setting is the extremality property of Gaussian states, which essentially asserts that the supremum of $\chi(Y;E)$ in Eqn.~\eqref{eqn:DW} is upper bounded by the value of $\chi(Y;E)$ computed for the Gaussian state $\rho_{AYE}^G$ with the same covariance matrix as $\rho_{AYE}$ \cite{GC06,NGA06}. In other words, it is bounded by a function that only depends on the covariance matrix of $\rho_{AYE}$, and even on the covariance matrix of $\rho_{AB}$ since the map $\mathcal{M}_{B \to Y}$ is fixed by the protocol and $\rho_{ABE}$ is an arbitrary purification of $\rho_{AB}$. 
The covariance matrix of $\rho_{AB}$ is defined as
\begin{align*}
\Gamma := \begin{bmatrix}
 \langle \hat{x}_A^2\rangle_\rho &\frac{1}{2} \langle \{\hat{x}_A, \hat{p}_A \}\rangle_\rho  &\frac{1}{2} \langle \{\hat{x}_A, \hat{x}_B \} \rangle_\rho &\frac{1}{2} \langle \{\hat{x}_A, \hat{p}_B\}\rangle_\rho\\
\frac{1}{2}\langle \{\hat{p}_A, \hat{x}_A \}\rangle_\rho & \langle  \hat{p}_A^2 \rangle_\rho  & \frac{1}{2}\langle \{ \hat{p}_A, \hat{x}_B \} \rangle_\rho & \frac{1}{2}\langle \{\hat{p}_A, \hat{p}_B\}\rangle_\rho\\
\frac{1}{2} \langle \{ \hat{x}_A,\hat{x}_B\} \rangle_\rho & \frac{1}{2}\langle \{\hat{x}_B, \hat{p}_A \}\rangle_\rho  & \langle  \hat{x}_B^2 \rangle_\rho & \frac{1}{2}\langle \{\hat{x}_B, \hat{p}_B\}\rangle_\rho\\
\frac{1}{2} \langle \{ \hat{p}_B,\hat{x}_A\} \rangle_\rho &\frac{1}{2} \langle \{\hat{p}_B, \hat{p}_A \}\rangle_\rho  &\frac{1}{2} \langle \{ \hat{p}_B, \hat{x}_B\} \rangle_\rho &  \langle \hat{p}_B^2\}\rangle_\rho
 \end{bmatrix}
\end{align*}
where we assume again without loss of generality that the first moment of the displacement operator vanishes.

Symmetry arguments (see \textit{e.g.} Appendix D of Ref.~\cite{lev15}) show that $\Gamma$ can be safely replaced by $\Gamma'$ when computing the secret key rate, with 
\begin{align*}
\Gamma' := \begin{bmatrix}
V \1_2 & Z \sigma_Z\\
Z \sigma_Z & W \1_2
\end{bmatrix}
\end{align*}
where the real numbers $V, W, Z$ are given by
\begin{align*}
V & :=\frac{1}{2}( \langle \hat{x}_A^2\rangle_\rho + \langle \hat{p}_A^2\rangle_\rho )= 1 + 2\tr( \rho \hat{a}^\dag \hat{a}),\\
W & := \frac{1}{2}(\langle \hat{x}_B^2\rangle_\rho + \langle \hat{p}_B^2\rangle_\rho)= 1 + 2\tr( \rho \hat{b}^\dag \hat{b}), \\
Z & := \frac{1}{4} \Big(  \langle \{\hat{x}_A, \hat{x}_B \} \rangle_\rho -  \langle \{\hat{p}_A, \hat{p}_B \} \rangle_\rho\Big) = \tr(\rho (\hat{a} \hat{b} + \hat{a}^\dag\hat{b}^\dag)),
\end{align*}
and $\sigma_Z$ is the Pauli matrix $\mathrm{diag}(1,-1)$. 
The Holevo information $\chi(Y;E)$ computed for the Gaussian state with covariance matrix $\Gamma'$ is given by
\begin{align}\label{eqn:chiYE}
\chi(Y;E) &= g\left(\frac{\nu_1-1}{2}\right)+ g\left(\frac{\nu_2-1}{2}\right)- g\left(\frac{\nu_3-1}{2}\right),
\end{align}
where $g(x) := (x+1) \log_2 (x+1) - x \log_2(x)$, $\nu_1$ and $\nu_2$ are the symplectic eigenvalues of $\Gamma'$ and $\nu_3$ depends on the choice of measurement setting (homodyne or heterodyne). The value of $\nu_3$ is given by $\nu_3 = V- \frac{Z^2}{W+1}$ in the heterodyne case and $\nu_3 = \sqrt{V(V- \frac{Z^2}{W})}$ in the homodyne case \cite{WPG12}. 

We note that both $X$ and $Y$ correspond to the expectations of local observables, namely $1 + 2 \hat{a}^\dag \hat{a}$ and $1 + 2 \hat{b}^\dag \hat{b}$. In particular, $X$ is simply a parameter of the protocol, which is independent of the quantum channel between Alice and Bob. It is customary in the literature to write it as 
\[ V = V_A+1,\]
where $V_A$ stands for the modulation variance. In general, this parameter can be optimized so as to maximize the secret key rate in a given experiment. For protocols with a Gaussian modulation, it is known that the optimal value of $V_A$ becomes larger and larger as the reconciliation efficiency $\beta$ gets closer and closer to 1. For discrete modulation schemes, such as the QPSK modulation, the optimal value of $V_A$ is much lower, and can even be significantly lower than the shot noise with current security proofs \cite{GGD19, LUL19}. 
The expectation $W$ is not fixed by the protocol, but can be measured locally by Bob who performs a heterodyne detection. 
The remaining quantity, $Z := \tr(\rho\, C)$ with 
\begin{align}\label{eqn:C}
C :=\hat{a} \hat{b} + \hat{a}^\dag\hat{b}^\dag,
\end{align}
will be the central object in the present work. If it could be measured directly in the protocol, then Alice and Bob would know the covariance matrix $\Gamma'$ and immediately get a bound on Eve's information. 
In particular, in any EB protocol, it is sufficient for Alice and Bob to both perform coherent measurements (homodyne or heterodyne) to obtain the covariance matrix. The security of such protocols is therefore well understood.
Unfortunately, these EB protocols are much less practical than PM protocols with a discrete modulation of coherent states, since they require the preparation of entangled states. For PM protocols, the state $\rho_{AB}$ does not actually exist in the lab. It is simply a convenient mathematical object, allowing us to discuss the security of the protocol. Consequently, it is in general impossible to infer what value $Z$ Alice and Bob would obtain if they really had access to $\rho_{AB}$. It is therefore necessary to find some indirect approach in order to get some bounds on $Z = \tr(\rho\, C)$.

Protocols with a Gaussian modulation (of Gaussian states) are an exception: in this case, one can easily compute this covariance matrix, and in particular the value of $Z=\tr(\rho\, C)$ from the data observed in the PM protocol \cite{GCW03}. The reason for this is that the measurement performed by Alice in the EB protocol is a Gaussian measurement, and therefore the observed statistics are sufficient to infer the covariance matrix. This is no longer the case for schemes with a discrete modulation: in that case, Alice performs a non-Gaussian measurement on the mode $A$ of $\rho_{AB}$ and this is in general insufficient to deduce the value of $\tr(\rho\, C)$, except by restricting the class of considered attacks \cite{LG09,LG11}. The main result of Ref.~\cite{GGD19} was to show that even if the exact value of $\tr(\rho\, C)$ cannot be recovered, it is still possible to obtain some bounds on this quantity by expressing it as the objective function of a semidefinite program.

%%%%%%%%%%%%%%%%%%%%%%%%%%%%%%%%%%%%%%%%
\section{Definition of the SDP and explicit solution}
\label{sec:def-sdp}

Our first goal is to specify the SDP we want to solve. 
As mentioned, the objective function is simply $\tr( \rho\, C)$ where $\rho_{AB}$ is the state shared by Alice and Bob, before they measure it, in the EB version of the protocol. In order to get the tightest possible bounds on the value of $\tr(\rho\, C)$, we need to impose some constraints on the possible states $\rho_{AB}$ that should be considered. These constraints have two origins: a first constraint merely says that $\rho_{AB}$ is obtained by applying some channel $\mathcal{N}_{A'\to B}$ to $|\Phi\rangle_{AA'}$; the other constraints come from observations made during the parameter estimation phase of the PM protocol. 

The first constraint turns out to be 
\begin{align}\label{eqn:constraint1}
\tr_B(\rho) = \bar{\tau},
\end{align}
which results from the fact that 
\[\tr_B(\rho) = \tr_B ( (\mathrm{id}_A \otimes \mathcal{N}_{A' \to B})( |\Phi\rangle \langle \Phi|)_{AA'} ) = \tr_{A'}(|\Phi\rangle \langle \Phi|) = \bar{\tau},\]
where we define $\bar{\tau}$ to be the complex conjugate of $\tau$ in the Fock basis. The choice of $\bar{\tau}$ may appear arbitrary at the moment, but will become clearer once we explain how to choose the purification $|\Phi\rangle$.
For the remaining constraints, we recall that Alice sends coherent states $|\alpha_k\rangle$ to Bob, and that they can gather information about the statistics corresponding to each such coherent state. 
Obviously, these statistics will need to be estimated properly during the protocol and one should endeavor to reduce the number of independent quantities that need to be estimated, since this number will greatly impact the key rate when taking finite-size effects into account.
The results that are readily available in the PM protocol are the first and second moments of the state received by Bob when Alice has sent $|\alpha_k\rangle$: 
\[ \beta_k := \tr(\rho_k b) \in \mathbbm{C}, \]
where $\rho_k := \mathcal{N}(|\alpha_k\rangle \langle \alpha_k|)$, as well as the second moment of Bob's state
\[ n_B := \tr(\rho \, b^\dag b).\]
Indeed, let us assume that a random sample of the measurement results of Bob when Alice sent the state $|\alpha_k\rangle$ are $\beta_{k,1}, \ldots, \beta_{k,N}$, then we expect that 
\[ \frac{1}{N} \sum_i \beta_{k,i} \xrightarrow[N\to\infty]{}  \tr(\rho_k b), \qquad  \frac{1}{N} \sum_{k,i} p_k |\beta_{k,i}|^2 \xrightarrow[N\to\infty]{} n_B+1.\] 
Recall that we consider collective attacks here, which means that the state $\rho_k$ is always the same (but unknown).
Bounding the speed of convergence of these empirical values is not completely trivial since we do not want to assume anything about the distribution of the $\beta_{k,i}$ but techniques similar to those developed in Ref.~\cite{lev15} can probably solve this issue. In any case, we do not worry about this specific difficulty here since we focus on asymptotic results and therefore assume that Alice and Bob are able to perform the parameter estimation step.

As mentioned, we ultimately wish to aggregate such values and only keep a few numbers, much less than $M$. 
Let us first relate these values to the bipartite state $\rho_{AB}$.
Without loss of generality, let us write 
\[ |\Phi\rangle = \sum_{k=1}^M \sqrt{p_k} |\psi_k\rangle |\alpha_k\rangle,\]
where the $\{|\psi_k\rangle\}$ form an orthonormal basis (that we will carefully choose later). 
With this notation, we obtain
\[ p_k \beta_k = \tr \Big(\rho (|\psi_k\rangle \langle \psi_k| \otimes \hat{b}) \Big).\] 
The second moment constraint is the easier one to deal with: we simply define the operator $\Pi \otimes b^\dag b$ where $\Pi := \sum_{k} |\psi_k\rangle\langle \psi_k|$ is a projector and observe that
\begin{align}\label{eqn:constraint2}
\tr( \rho (\Pi \otimes b^\dag b)) = n_B, 
\end{align}
where the right-hand side can be measured in the protocol. 
In order to define the first moment constraints, we need to introduce an operator that will play a central role in our analysis:
\begin{align}
a_\tau := {\tau}^{1/2} a {\tau}^{-1/2}. \label{eqn:atau}
\end{align}
We will rely on two first-moment constraints:
\begin{align}\label{eqn:constraint3}
\tr\Big(\rho \, C_1 \Big) =2 c_1, \qquad 
\tr\Big(\rho\, C_2 \Big) = 2 c_2,
\end{align}
with operators $C_1$ and $C_2$ defined by 
\begin{align}\label{eqn:C1}
C_1 := \sum_k  \overline{\langle {\alpha_k} |a_\tau|\alpha_k\rangle}  |\psi_k\rangle \langle \psi_k| \otimes \hat{b} +\mathrm{h.c.}, \quad C_2 := \sum_k  \bar{ \alpha_k}   |\psi_k\rangle \langle \psi_k| \otimes \hat{b} +\mathrm{h.c.}
\end{align}
The correlation coefficients $c_1$ and $c_2$ can be estimated experimentally by
\[ c_1 = \mathrm{Re}\Big(\sum_k p_k \overline{ \langle \alpha_k |a_\tau|\alpha_k\rangle} \beta_k\Big), \qquad c_2 = \mathrm{Re}\Big( \sum_k   p_k \bar{\alpha_k} \beta_k \Big).\]
Here, h.c.~stands for Hermitian conjugate, and we use $\bar{\cdot}$ to denote the complex conjugation (with respect to the Fock basis).
If we introduce the vectors ${\bm \alpha} := (\alpha_k)_k, {\bm \alpha_\tau} := ( \langle \alpha_k |a_\tau |\alpha_k\rangle)_k$ and ${\bm \beta} = (\beta_k)_k$, then the values of $c_1$ and $c_2$ are simply the following inner products:
\[ c_1 = \mathrm{Re}( {\bm \alpha_\tau} | {\bm\beta}), \qquad c_2 = \mathrm{Re}({\bm\alpha} | {\bm \beta}),\]
where we define the weighted inner product $({\bm x} |{\bm y}) := \sum p_k \bar{x_k} y_k$. 
Of course, the specific form of the operator $C_1$ may look somewhat mysterious at this point since it is not clear why the operator $\hat{a}_\tau = \tau^{1/2} \hat{a} \tau^{-1/2}$ should play any role at all in the problem, and why $c_1$ should be a meaningful quantity to estimate during the protocol. The story goes in the other direction: the constraints that should be monitored during the PM protocol are clearly functions of the $\beta_{k}$'s, since they are the only observable values in the PM protocol. The simplest such constraints are linear functions in the moments of $\beta_{k}$ and since our proofs will ultimately rely on the extremality properties of the Gaussian states, it makes sense to focus on the first and second moments\footnote{We also tried to add fourth moment constraints, similarly to Ref.~\cite{LUL19}, for the QPSK modulation but this did not significantly improve the performance. In addition, it is not clear how to obtain analytical bounds that exploit such constraints, and it is important to recall that any such constraint leads to a quantity that needs to be estimated experimentally, and that will contribute to finite-size effects. Overall, it thus seems much easier to focus exclusively on the first two moments of the quantum state.}. The relevant second moment is the variance of $\beta_k$, but there is no obvious candidate for the first moment conditions. Our strategy was therefore to optimize the first moment conditions by leaving them as general as possible and only later pick the relevant ones. This is exactly how we arrived at the definitions of $C_1$ and $C_2$.

The constraints of Eqn.~\eqref{eqn:constraint1}, \eqref{eqn:constraint2} and \eqref{eqn:constraint3} are the only ones we will impose in addition to $\rho \succeq 0$. 
Since the secret key rate is minimized when the value of $Z = \tr(\rho\, C)$ is minimal\footnote{We do not have a formal proof of this claim but have checked it numerically. In any case, for given parameters, one should consider the maximum of $\chi(Y;E)$ for $Z$ in the interval given by Eqn.~\eqref{eqn:main-result}.}, we finally state our main SDP:
\begin{align}
\min \quad & \tr(\rho\, C) \label{sdp:primal}\\
\mathrm{s.t.} \quad  &\left\{   \begin{array}{l}\tr_B(\rho) = \bar{\tau} \nonumber\\
\tr\Big(\rho \Big( \sum_k \overline{ \langle \alpha_k |a_\tau|\alpha_k\rangle}  |\psi_k\rangle \langle \psi_k| \otimes \hat{b} +\mathrm{h.c.}\Big) \Big) =2 c_1 \nonumber\\
\tr\Big(\rho \Big( \sum_k   \bar{\alpha_k}   |\psi_k\rangle \langle \psi_k| \otimes \hat{b} +\mathrm{h.c.}\Big) \Big) = 2 c_2, \nonumber\\
\tr( \rho (\Pi \otimes \hat{b}^\dag \hat{b})) = n_B,\nonumber\\
\rho \succeq 0.
\end{array}\right.
\end{align}
Our main technical contribution is to provide the following bounds for the interval of possible values for $\tr(\rho \, C)$ under these constraints:
\begin{align}\label{eqn:main-result}
\tr(\rho \, C) \in \left[  2 c_1 - 2\sqrt{w \Big(n_B - \frac{c_2^2}{\langle n \rangle}\Big) },  2 c_1 + 2\sqrt{ w \Big(n_B - \frac{c_2^2}{\langle n \rangle}\Big) }\right],
\end{align}
where we recall that $\langle n \rangle = \sum_k p_k |\alpha_k|^2$ is the average photon number in the modulation and we define the quantity
\begin{align}\label{eqn:W}
w:=  \sum_k p_k \left( \langle \alpha_k | a_\tau^\dag a_\tau |\alpha_k\rangle - |\langle \alpha_k | a_\tau |\alpha_k\rangle|^2\right).
\end{align}
The Cauchy-Schwarz inequality, $| ({\bm \alpha}| {\bm\beta}) |^2 \leq  ({\bm \alpha}| {\bm\alpha})({\bm \beta}| {\bm\beta}) $, implies that the term $n_B - \frac{c_2^2}{\langle n\rangle}$ is nonnegative since $\langle n \rangle = ({\bm\alpha} | {\bm \alpha})$, $c_2 = \mathrm{Re}({\bm\alpha} | {\bm \beta})$ and $n_B \geq ({\bm\beta} | {\bm \beta})$ (with equality when $\rho_k = |\beta_k\rangle\langle \beta_k|)$. The quantitiy $n_B - \frac{c_2^2}{\langle n\rangle}$ is (proportional to) the excess noise, corresponding to the noise added by the quantum channel.
Here, both $\langle n \rangle$ and $chi$ are fixed by the choice of the constellation. 
In particular, inserting the lower bound 
\begin{align}\label{eqn:Z^*}
Z^* :=2 c_1 - 2\left(\Big(n_B - \frac{c_2^2}{\langle n \rangle}\Big) w \right)^{1/2}
\end{align}
of the interval in the covariance matrix $\Gamma'$ and computing the associated Holevo bound yields an analytical lower bound on the asymptotic secret key rate of the CV QKD protocol\footnote{Note that while the minimum value in the interval of Eqn.~\eqref{eqn:main-result} yields the maximum value of the Holevo information defined in Eqn.~\eqref{eqn:chiYE} in most cases, in all generality, one should simply consider the value of the interval that maximizes the Holevo information.}.

We note that an important feature of $Z^*$ is that it only involves 3 quantities that need to be determined experimentally. In particular, there is no need for the precise knowledge of all the $\beta_k$, which would make any finite-size analysis very challenging. At the same time, $c_1$ is an additional quantity that was not present in previous works, for instance in the definition of the SDP in Ref.~\cite{GGD19}. While this difference does not appear in simulations of a Gaussian quantum channel since the ratio between $c_1$ and $c_2$ is fixed in that case, it does play a role in a real experiment, and will also impact the finite-size secret key rate since an additional parameter needs to be estimated.

As we discuss in more details in Section \ref{sec:Gaussian}, a simple calculation shows that $a_{\tau_{\mathrm{G}}} = \sqrt{\frac{1+\langle n \rangle}{\langle n \rangle}} \hat{a}$ and therefore $w=0$ in the Gaussian case, recovering the well-known result that the covariance term is completely determined, and hence does not depend on the excess noise, for a Gaussian modulation. In particular, there are only two independent experimental quantities to monitor in that case, $c_1$ and $n_B$.

\paragraph{Expected bound for a Gaussian quantum channel.}
The bound of Eqn.~\eqref{eqn:main-result} can be readily used in any experimental implementation of the protocol, but it is also useful to be able to get an estimate of such a bound for a typical experimental setup. 
In particular, since most experiments are implemented in fiber, it is typical to model the expected quantum channel between Alice and Bob as a phase insensitive Gaussian channel characterized by a transmittance $T$ and an excess noise $\xi$. 
This means that if the input state is a coherent state $|\alpha\rangle$, then the output state is a displaced thermal state centered at $\sqrt{T} \alpha$ with a variance given by $1+T\xi$. In other words, the random variable $\beta_k$ can be modeled as 
\[ \beta_k = \sqrt{T} \alpha_k + \gamma_k,\]
where $\gamma_k$ is a Gaussian random variable corresponding to the shot noise (of variance 1 with our choice of units) and to the excess noise (of variance $T\xi$).
In this case, one can readily compute the expected values of $c_1$, $c_2$ and $n_B$ (see Section \ref{sec:primal} for details):
\begin{align*}
c_1 &= \sqrt{T} \; \tr(\bar{\tau}^{1/2} a \bar{\tau}^{1/2} a^\dag) \\
c_2 &= \sqrt{T} \langle n \rangle,\\
n_B &= T \langle n \rangle + T \frac{\xi}{2},
\end{align*}
which yields a minimum value $Z^*(T,\xi) = \min \tr(\rho \, C)$ equal to
\begin{align}\label{eqn:bosonic}
Z^*(T,\xi) = 2\sqrt{T} \; \tr(\tau^{1/2} a \tau^{1/2} a^\dag) -  \sqrt{2T\xi w}.
\end{align}
The linear dependence in $\sqrt{T}$ is expected, and we note that the correction term, scaling like $\sqrt{\xi}$, heavily impacts the value of the covariance, for nonzero excess noise, unless $w$ is very small. As we will later see, while $W$ is rather large and leads to rather poor performance in the case of a QPSK modulation with only four coherent states, this is no longer the case for larger constellations, for instance with a 64-QAM of 64 coherent states. 
Eqn.~\eqref{eqn:bosonic} is generalized to the case of a modulation of arbitrary states in Eqns \eqref{eqn:final-upp} and \eqref{eqn:final-low}.

%%%%%%%%%%%%%%%%%%%%%%%%%%%%%%%%%%%%%%%%
\section{Analytical study of the SDP}
\label{sec:primal}

In this section, we detail how to obtain a lower bound on the value of the primal SDP of Eqn.~\eqref{sdp:primal}.
In fact, although it is primarily the minimum of the objective function that is relevant for CV QKD, we can  more generally aim to find the whole interval of values for $\tr(\rho\, C)$ compatible with the constraints.
We start by explaining how to choose a convenient purification of $\tau$ and how to model Alice's measurement in the entanglement-based version of the protocol and then proceed to obtain our main result.

\subsection{Purification of $\tau$}

Before proceeding with the change of variables, let us discuss the choice of the purification $|\Phi\rangle$ for the modulation state $\tau$. We choose 
\begin{align}\label{eqn:Phi}
|\Phi\rangle := (\1 \otimes \tau^{1/2}) \sum_{n=0}^\infty |n\rangle |n\rangle.
\end{align}
By writing the spectral decomposition of $\tau$:
\[ \tau = \sum_{k=1}^M \lambda_k |\phi_k\rangle \langle \phi_k|,\]
we immediately obtain 
\[ |\Phi\rangle = \sum_{k=1}^M \lambda_k^{1/2}  |\bar{\phi_k}\rangle  |\phi_k\rangle,\]
where $|\bar{\phi_k}\rangle$ is obtained by conjugating the coefficients of $|\phi_k\rangle$ in the Fock basis. 
Note that we can also write\footnote{In an earlier version of this paper, see \cite{DBL21}, we restricted the analysis to constellations which are symmetric under complex conjugation, in the sense that the coherent states $|\alpha_k\rangle$ and $|\bar{\alpha_k}\rangle$ are sent with the same probability. This is essentially without loss of generality since all reasonable constellations used in telecommunications satisfy this property. The main advantage is some slight simplification of the formula since we could use $\bar{\tau} = \tau$ everywhere. However, it is useful to relax this constraint if one wants to study possible imperfections in the state preparation of the protocol for instance.} $|\Phi\rangle = (\bar{\tau}^{1/2} \otimes \1) \sum_{n=0}^\infty |n\rangle |n\rangle$.
Considering $\bar{\tau}^{-1/2}$ to be the square-root of the Moore-Penrose pseudo-inverse of $\bar{\tau}$, equal to the inverse of $\bar{\tau}$ on its support and to zero elsewhere (recall that $\bar{\tau} = \sum_{k=1}^M p_k |\bar{\alpha_k}\rangle \langle \bar{\alpha_k}|$ is an operator of rank $M$ since any finite set of coherent states forms an independent family), we have that
\[ (\bar{\tau}^{-1/2} \otimes \1) |\Phi\rangle = (\Pi \otimes \1) \sum_{n=0}^\infty |n\rangle |n\rangle = \sum_{k=1}^M |\bar{\phi_k}\rangle |\phi_k\rangle,\]
where $\Pi = \sum_{k=1}^M |\bar{\phi_k}\rangle \langle \bar{\phi_k}|$ is the orthogonal projector onto the $M$-dimensional subspace spanned by the (conjugated) coherent states $|\bar{\alpha_k}\rangle$ of the modulation (equivalently, $\Pi$ is the projector onto the support of $\bar{\tau}$).
Note indeed that the $|\phi_k\rangle$ (as well as the $|\bar{\phi_k}\rangle$) are orthogonal since they appear in the spectral decomposition of $\tau$. This means that $(\bar{\tau}^{-1/2} \otimes \1) |\Phi\rangle$ is an $M$-dimensional maximally entangled state. 
We define the state $|\psi_k\rangle$ by\footnote{This definition should be modified for protocols relying on a modulation of thermal states $\tau_k$, as mentioned in Section \ref{sec:thermal} for instance. In that case, one would define operators of the form $p_k \bar{\tau}^{-1/2} \bar{\tau_k} \bar{\tau}^{-1/2}$.}
\begin{align}\label{eqn:psi}
|\psi_k\rangle := \sqrt{p_k} \bar{\tau}^{-1/2} |\bar{\alpha_k}\rangle.
\end{align}
Note that
\[ \sum_{k=1}^M |\psi_k\rangle\langle \psi_k| = \sum_{k=1}^M p_k \bar{\tau}^{-1/2}  |\bar{\alpha_k}\rangle\langle \bar{\alpha_k}| \bar{\tau}^{-1/2} = \bar{\tau}^{-1/2} \bar{\tau} \bar{\tau}^{-1/2} = \Pi.\]
From this, we conclude that the family $\{|\psi_k\rangle\}$ forms an orthonormal basis for the relevant subspace, and moreover, we obtain\footnote{To see this, we can simply compute the overlap between this state and the definition $(\bar{\tau}^{1/2} \otimes \1)\sum_n |n\rangle |n\rangle$:
\begin{align*} \sum_k \sqrt{p_k} \langle \psi_k|\langle \alpha_k| (\bar{\tau}^{1/2} \otimes \1)\sum_n |n\rangle |n\rangle &= \sum_k p_k \langle \bar{\alpha_k}| \langle \alpha_k |(\bar{\tau}^{-1/2} \otimes \1) (\bar{\tau}^{1/2} \otimes \1) )\sum_n |n\rangle |n\rangle\\
&= \sum_k p_k \langle \bar{\alpha_k}| \Pi |\bar{\alpha_k}\rangle = 1,
\end{align*}
where we used that $\langle \alpha_k| \sum_n |n\rangle |n\rangle= |\bar{\alpha_k}\rangle$ and $\bar{\tau}^{-1/2} \bar{\tau}^{1/2} =\Pi$.
}
\begin{align}\label{eqn:Phi2}
|\Phi\rangle = \sum_{k=1}^M \sqrt{p_k} |\psi_k\rangle |\alpha_k\rangle.
\end{align}
An interpretation of the states $|\psi_k\rangle$ is that they define the projective measurement that Alice should perform in the entanglement-based version of the protocol in order to recover the Prepare-and-Measure protocol: if Alice measures her state and obtains the result indexed by $k$, then the second mode of $|\Phi\rangle$, the one which is sent through the quantum channel to Bob, collapses to $|\alpha_k\rangle$.

\subsection{The Sum-Of-Squares}
\label{sec:sos}
Now that we have defined the states $|\psi_k\rangle$, we are ready to analyze the SDP of Eqn.~\eqref{sdp:primal}, which we recall here for convenience:
\begin{align*}
\min \quad & \tr(\rho\, C) \\
\mathrm{s.t.} \quad  &\left\{   \begin{array}{l}\tr_B(\rho) = \bar{ \tau} \nonumber\\
\tr\Big(\rho \, C_1 \Big) =2 c_1 \nonumber\\
\tr\Big(\rho \, C_2 \Big) = 2 c_2, \nonumber\\
\tr( \rho (\Pi \otimes b^\dag b)) = n_B,\nonumber\\
\rho \succeq 0,
\end{array}\right.
\end{align*}
with $C = ab + a^\dag b^\dag$, $C_1 = \sum_k  \overline{\langle {\alpha_k} |a_\tau|\alpha_k\rangle}  |\psi_k\rangle \langle \psi_k| \otimes \hat{b} +\mathrm{h.c.}$ and $C_2 = \sum_k  \bar{ \alpha_k}   |\psi_k\rangle \langle \psi_k| \otimes \hat{b} +\mathrm{h.c.}$

In order to get explicit bounds on $\tr(\rho\, C)$ for feasible points of this program, we exploit a standard technique called sum-of-squares. It consists in exhibiting some clever nonnegative operator (namely $KK^\dag$ below) such that we can bound the value of $\tr\big(\rho(C- KK^\dag )\big)$ from the constraints of the program. In that case, we immediately get
\[ \tr(\rho\, C) = \tr(\rho ( C- KK^\dag)) + \tr(\rho KK^\dag) \geq \tr(\rho (C - KK^\dag)),\]
where we used that $\tr(\rho KK^\dag) \geq 0$. Finding an operator $K$ that will give a good bound on the value of the SDP is nontrivial, and the problem is even more complicated here because the relevant operators live in an infinite-dimensional Hilbert space. 
In a previous version of this manuscript (Ref.~\cite{DBL21}), we attacked this problem by first performing a change of variables consisting in displacing Bob's system by $-t\alpha_k$ (for an optimized value of $t$) when the state prepared by Alice is $|\alpha_k\rangle$. The advantage of this procedure was that the new state held by Bob has a very low average photon number and is therefore close to the vacuum state (and equal to it when there is no excess noise). It was then possible to guess what would be a good parameterized sum-of-squares. In the present version of the manuscript, we bypass this change-of-variable altogether and directly define the relevant operators:
\begin{align*} 
A &:= \Pi a \Pi,\\
B &:= \sum_k |\psi_k\rangle \langle \psi_k| \otimes (b - t\alpha_k)\\
P &:= \sum_k y_k |\psi_k\rangle \langle \psi_k|, \\
K_{\pm} &:= z(A-xP^\dag) \pm \frac{1}{z} B^\dag,
\end{align*}
where the scalars $t$, $\{ y_k\}_k$, $x$ and $z$ will be optimized later. The proof ends up being much shorter, involving fewer algebraic operations, but may seem a bit magical.
From $KK^\dag \succeq 0$, we infer that $\tr(\rho KK^\dag) \geq 0$. Expanding this expression, we find
\begin{align} \label{eqn:sos}
 K_{\pm} K_{\pm}^\dag =z^2(A- xP^\dag)(A^\dag - xP) \pm (AB + B^\dag A^\dag) \mp x(P^\dag B + B^\dag P) + \frac{1}{z^2} B^\dag B.
\end{align}
Let us consider each of these four terms individually and take their expectation with respect to the state $\rho$.

\begin{enumerate}
\item The first term 
\[ z^2 \, \tr( \rho (A- xP^\dag)(A^\dag - xP))\]
is a quadratic form in $x$, and is minimal for the choice 
\begin{align}\label{eqn:x}
x = \frac{1}{2} \frac{ \tr(\rho(AP + P^\dag A^\dag))}{\tr(\rho P^\dag P)}.
\end{align}
We then get
\begin{align*}
\tr( \rho (A- xP^\dag)(A^\dag - xP)) &= \tr(\rho A A^\dag) - \frac{1}{4}\frac{ \Big(\tr(\rho(AP + P^\dag A^\dag)) \Big)^2}{\tr(\rho P^\dag P)}\\
&= \tr(\bar{\tau}  a \Pi a^\dag) - \frac{ \Big(\mathrm{Re}(\tr(\bar{\tau} AP))\Big)^2}{\tr(\bar{\tau} P^\dag P)},
\end{align*}
where we exploited in the second line the fact that the operators $A $ and $P$ all act on the first subsystem and $\tr_{B}(\rho) = \bar{\tau}$.
Recalling the definition of the operator $a_{\bar{\tau}} := \bar{\tau}^{1/2} a\bar{\tau}^{-1/2}$, the first term becomes
\[  \tr(\bar{\tau}  a \Pi a^\dag) =\tr( \bar{\tau} a_{\bar{\tau}}^\dag a_{\bar{\tau}}).\]
We will optimize the choice of $P$ in order to maximize the fraction $\frac{\Big(\mathrm{Re}(\tr(\bar{\tau} AP))\Big)^2}{\tr(\bar{\tau} P^\dag P)}$. By definition of $P$, we have:
\[ \tr(\bar{\tau} a P) = \sum_k y_k \langle \psi_k |\bar{\tau} a|\psi_k\rangle, \qquad \tr(\bar{\tau} P^\dag P) = \sum_k |y_k|^2 \langle \psi_k |\bar{\tau} |\psi_k\rangle.\]
We write $y_k = x_k e^{i\theta_k}$ with $x_k\geq 0$ and choose $\theta_k$ so that $y_k \langle \psi_k |\bar{\tau} a|\psi_k\rangle = x_k |\langle \psi_k |\bar{\tau} a|\psi_k\rangle|$ is nonnegative.
We obtain
\begin{align*}
\frac{\Big(\mathrm{Re}(\tr(\bar{\tau} AP))\Big)^2}{\tr(\bar{\tau} P^\dag P)}&= \frac {\Big(  \sum_k x_k | \langle \psi_k |\bar{\tau} a|\psi_k\rangle|\Big)^2}{\sum_k x_k^2 \langle \psi_k |\bar{\tau} |\psi_k\rangle}.
\end{align*}
Choosing $x_k := \frac{ | \langle \psi_k |\bar{\tau} a|\psi_k\rangle|}{ \langle \psi_k |\bar{\tau} |\psi_k\rangle}$, we get
\begin{align*}
\frac{\Big(\mathrm{Re}(\tr(\bar{\tau} AP))\Big)^2}{\tr(\bar{\tau} P^\dag P)} &=  \sum_k  \frac { | \langle \psi_k |\bar{\tau} a|\psi_k\rangle|^2}{\langle \psi_k |\bar{\tau} |\psi_k\rangle} = \sum_k p_k |\langle \bar{\alpha_k} |\bar{\tau}^{1/2} a \bar{\tau}^{-1/2} |\bar{\alpha_k}\rangle|^2 = \sum_k p_k |\langle \bar{\alpha_k} |a_{\bar{\tau}} |\bar{\alpha_k}\rangle|^2,
\end{align*}
where we exploited that $|\psi_k\rangle = \sqrt{p_k} \bar{\tau}^{-1/2}|\bar{\alpha_k}\rangle$ in the second equality. 
We finally obtain for our choice of $x$ and $P$ that 
\begin{align}\label{eqn:term1}
 z^2 \, \tr( \rho (A- xP^\dag)(A^\dag - xP)) = z^2 \big( \tr( {\tau} a_{{\tau}}^\dag a_{{\tau}}) -\sum_k p_k |\langle {\alpha_k} |a_{{\tau}} |{\alpha_k}\rangle|^2  \big),
\end{align}
where we exploited the invariance of the expression under complex conjugation.
This equals $z^2 w$ with $w$ defined in Eqn.~\eqref{eqn:W}.

\item We turn to the second term of Eqn.~\eqref{eqn:sos}. By definition, 
\[ AB = \sum_k \Pi a |\psi_k\rangle \langle \psi_k| \otimes (b - t\alpha_k),\]
and therefore
\begin{align*}
\tr( \rho (AB + B^\dag A^\dag) ) &= \tr( \rho (Ab + A^\dag b^\dag) ) - t \, \tr\Big(\rho\big( \sum_{k,\ell}  \alpha_k \langle \psi_\ell|a|\psi_k\rangle   |\psi_\ell \rangle \langle \psi_k | + \mathrm{h.c.} \big) \Big)\\ 
&= \tr(\rho \, C) - t \, \Big( \sum_{k,\ell} \sqrt{p_k p_\ell} \alpha_k \langle \psi_\ell|a|\psi_k\rangle  \tr(  |{\alpha}_k \rangle \langle {\alpha}_\ell | )+ \mathrm{c.c.}    \Big)\\
&= \tr(\rho\, C) - t \, \Big( \sum_{k,\ell} \sqrt{p_k p_\ell}  \langle \psi_\ell|a|\psi_k\rangle  \langle \alpha_\ell | b |\alpha_k\rangle+ \mathrm{c.c.}    \Big)\\
&= \tr(\rho \, C) - t \, \langle \Phi | ab + a^\dag b^\dag |\Phi\rangle.
\end{align*}
In the second line, $\mathrm{c.c.}$ stands for complex conjugate.
One can simplify the second term further and write it as a function of $\tau$.
From $(\1 \otimes \hat{b}) \sum_{n=0}^\infty |n\rangle |n\rangle = (\hat{a}^\dag \otimes \1)\sum_{n=0}^\infty |n\rangle |n\rangle$, we obtain
\begin{align*}
\langle \Phi | ab |\Phi\rangle &= \sum_{m,n=0}^\infty \langle m|\langle m | (\bar{\tau}^{1/2} \otimes \1) ab (\bar{\tau}^{1/2} \otimes \1)|n\rangle |n\rangle = \sum_{m,n=0}^\infty \langle m | \bar{\tau}^{1/2} a \bar{\tau}^{1/2}  a^\dag |n\rangle \langle m|n\rangle\\
& = \tr(\bar{\tau}^{1/2} a \bar{\tau}^{1/2}a^\dag).
\end{align*}
This expression is real since it equals the trace of the Hermitian matrix $ \bar{\tau}^{1/4} a \bar{\tau}^{1/2} a^\dag \bar{\tau}^{1/4}$. 
In particular, it is invariant under complex conjugation, and we finally get the following expression for the second term of Eqn.~\eqref{eqn:sos}:
\begin{align}\label{eqn:term2}
\tr( \rho (AB + B^\dag A^\dag) ) =  \tr(\rho\, C) - 2 t  \,  \tr(\bar{\tau}^{1/2} a \bar{\tau}^{1/2}a^\dag).
\end{align}

\item The third term of Eqn.~\eqref{eqn:sos} can be computed directly:
\begin{align*}
x \, \tr(\rho ( P^\dag B + B^\dag P))) &= x \, \tr \Big( \rho \big(\sum_k \bar{y_k} |\psi_k\rangle \langle \psi_k| \otimes (b- t\alpha_k) \big)\Big) + \mathrm{c.c.}\\
&= x \, \tr \Big( \sum_k p_k \bar{y_k} |\psi_k\rangle \langle \psi_k| \otimes \sum_r E_r |\alpha_k\rangle \langle \alpha_k| E_r^\dag (b- t\alpha_k) \Big) + \mathrm{c.c.}\\
&= x \sum_k p_k \bar{y_k} \langle \alpha_k | \sum_r E_r^\dag(b-t\alpha_k) E_r |\alpha_k\rangle + \mathrm{c.c.}
\end{align*}
where we introduced the Kraus operators $\{ E_r\}_r$ of the channel $\mathcal{N}: A' \to B$, which satisfy $\sum_r E_r^\dag E_r = \1$ in the second line and wrote the bipartite state $\rho$ as
\[ \rho = \sum_{k,\ell} \sqrt{p_k p_\ell} |\psi_k\rangle\langle \psi_\ell| \otimes \sum_r E_r |\alpha_k\rangle \langle \alpha_\ell| E_r^\dag.\]
With our choice of $\{y_k\}_k$, this simplifies to
\begin{align*}
x \, \tr(\rho ( P^\dag B + B^\dag P)) &= x\sum_{k,r} \langle \psi_k |\bar{\tau} a|\psi_k\rangle \langle \alpha_k |E_r^\dag(b-t \alpha_k) E_r| \alpha_k\rangle + \mathrm{c.c.}
\end{align*}
We also observe that $\tr(\bar{\tau} AP) = \tr(\bar{\tau} P^\dag P)$, which implies that our optimized value of $x$ equals 1. 
Recalling that $\langle \psi_k|\bar{\tau} a|\psi_k\rangle = p_k \langle \bar{\alpha_k}| a_{\bar{\tau}}| \bar{\alpha_k}\rangle$, this gives
\begin{align}
x \, \tr(\rho ( P^\dag B + B^\dag P)) &= \sum_k  p_k \langle \bar{\alpha_k}| a_{\bar{\tau}}| \bar{\alpha_k}\rangle \tr ( \sum_r E_r |\alpha_k\rangle \langle \alpha_k|E_r^\dag \,   b) -t \sum_k \alpha_k \langle \psi_k |\bar{\tau} a|\psi_k\rangle + \mathrm{c.c.} \nonumber\\
& = \tr\Big( \rho \big( \sum_k \langle \bar{\alpha_k}| a_{\bar{\tau}}| \bar{\alpha_k}\rangle |\psi_k\rangle \langle \psi_k| \otimes b + \mathrm{h.c.} \big)\Big) -t \,  \langle \Phi |ab + a^\dag b^\dag|\Phi\rangle \nonumber\\
&= 2 c_1 -2 t\, \tr( \tau^{1/2} a \tau^{1/2} a^\dag), \label{eqn:term3}
\end{align}
where we exploited the constraint $\tr(\rho\, C_1) = 2c_1$ in the last equality.

\item The final term of Eqn.~\eqref{eqn:sos} is 
\[ \frac{1}{z^2} \tr(\rho B^\dag B) = \frac{1}{z^2} \tr\Big(\rho \sum_k |\psi_k\rangle \langle \psi_k| \otimes ( b^\dag b - t( \alpha_k b^\dag + \bar{\alpha_k} b) + t^2 |\alpha_k|^2) \Big).\]
The three subterms give respectively $n_B$, $2tc_2$ and $t^2 \langle n\rangle$ (where $\langle n \rangle$ is the average photon number in the constellation). Overall this term becomes
\begin{align}\label{eqn:term4}
\frac{1}{z^2} \tr(\rho B^\dag B) = \frac{1}{z^2} (n_B - 2tc_2 +t^2 \langle n\rangle).
\end{align}

\end{enumerate}

Putting Eqn.~\eqref{eqn:term1}, \eqref{eqn:term2}, \eqref{eqn:term3} and \eqref{eqn:term4} together, we get that 
\begin{align*}
z^2 w \pm \Big(\tr(\rho \, C) - 2 t  \,  \tr(\bar{\tau}^{1/2} a \bar{\tau}^{1/2}a^\dag)\Big) \mp \Big( 2 c_1 -2 t\, \tr( \tau^{1/2} a \tau^{1/2} a^\dag) \Big) +  \frac{1}{z^2} (n_B - 2tc_2 +t^2 \langle n\rangle)
\end{align*}
is nonnegative, which is equivalent to 
\begin{align*}
\tr(\rho\, C) &\geq 2 c_1 - z^2 w - \frac{1}{z^2}  (n_B - 2tc_2 +t^2 \langle n\rangle),\\
\tr(\rho\, C) &\leq 2 c_1 + z^2 w + \frac{1}{z^2}  (n_B - 2tc_2 +t^2 \langle n\rangle).
\end{align*}

Optimizing over the variables $t$ and $z$ with\footnote{In some cases, for instance with a Gaussian modulation, the term $W$ corresponding to $z^2$ vanishes. One should then consider the limit $z \to \infty$ in the optimization below.}
\[t = \frac{c_2}{\langle n \rangle} \quad \text{and} \quad z^4 = \frac{ n_B - \frac{c_2^2}{\langle n \rangle}}{w},\]
we obtain
\[  2 c_1 - 2\left(\Big(n_B - \frac{c_2^2}{\langle n \rangle}\Big) w \right)^{1/2} \leq \tr(\rho \, C)\leq  2 c_1 + 2\left(\Big(n_B - \frac{c_2^2}{\langle n \rangle}\Big) w \right)^{1/2}.
\]
This concludes our proof.

%%%%%%%%%%%%%%%%%%%%%%%%%%%%%%%%%%%%%%%%
\section{The Gaussian modulation}
\label{sec:Gaussian}

In this section, we show that the formula from Eqn.~\eqref{eqn:bosonic} gives the standard value for a Gaussian modulation \cite{GC06}.
Let us consider a modulation such that $\tau_G$ has $\langle n \rangle$ photons on average:
\[ \tau_{\mathrm{G}} = \frac{1}{\pi \langle n \rangle} \int_{\mathbbm{C}} \exp\left( -\frac{1}{\langle n \rangle} |\alpha|^2\right) |\alpha\rangle \langle \alpha| d\alpha = \frac{1}{1+\langle n \rangle} \sum_{m=0}^\infty \left(\frac{\langle n \rangle}{1+\langle n \rangle}\right)^m |m\rangle \langle m|.\] 
Computing $a_{\tau_{\mathrm{G}}} = \tau_{\mathrm{G}} ^{1/2} a \tau_{\mathrm{G}}^{-1/2}$ is straightforward:
\begin{align*}
a_{\tau_{\mathrm{G}}} &= \sum_{m,n=0}^\infty \left(\frac{\langle n \rangle}{1+\langle n \rangle}\right)^{(m-n)/2} |m\rangle \langle m| a |n\rangle \langle n|\\
&= \sum_{m,n=0}^\infty \left(\frac{\langle n \rangle}{1+\langle n \rangle}\right)^{(m-n)/2} |m\rangle \langle n|   \sqrt{n} \langle m |n-1\rangle\\
&= \sum_{n=1}^\infty \left(\frac{\langle n \rangle}{1+\langle n \rangle}\right)^{-1/2} \sqrt{n} |n-1\rangle \langle n|   \\
&= \left(1 + \frac{1}{\langle n \rangle}\right)^{1/2} a 
\end{align*}
and we observe that it is simply a rescaling of the original annihilation operator. In particular, coherent states are eigenstates for $a_{\tau_{\mathrm{G}}}$ and we obtain
\[ \langle \alpha | a_{\tau_{\mathrm{G}}}^\dag a_{\tau_{\mathrm{G}}} |\alpha \rangle = \left(1 + \frac{1}{\langle n \rangle}\right) \langle \alpha | a^\dag a |\alpha \rangle = |\langle \alpha |a_{\tau_{\mathrm{G}}} |\alpha \rangle|^2,\]
which shows that $w$ vanishes for a Gaussian modulation.
This shows that 
\[ \tr(\rho \, C) = 2c_1\]
with 
\begin{align*}
c_1 =  \mathrm{Re}( {\bm \alpha_\tau} | {\bm\beta})
= \left(1 + \frac{1}{\langle n \rangle}\right)^{1/2}  \mathrm{Re}({\bm\alpha} | {\bm \beta}).
\end{align*}
In particular, if the transmittance of the channel is $T$, meaning that ${\bm \beta} = \sqrt{T} {\bm \alpha}$, we get $\mathrm{Re}({\bm\alpha} | {\bm \beta}) = \sqrt{T} \langle n \rangle$ and recover the standard value for a Gaussian modulation
\[ \tr(\rho\, C) = 2 \sqrt{T} \sqrt{\langle n \rangle^2 + \langle n \rangle}.\]

\paragraph{Interpretation of $w$.}
What is remarkable in the case of a Gaussian modulation is that the quantity $w$ vanishes. 
Note that $w$ is the expectation of 
\[  \langle \alpha_k | a_\tau^\dag a_\tau |\alpha_k\rangle - |\langle \alpha_k | a_\tau |\alpha_k\rangle|^2\]
and it vanishes here because each such term vanishes. This results from the fact that any coherent state $|\alpha\rangle$ is an eigenstate of the operator $\hat{a}_\tau$, which is simply a rescaled version of the annihilation operator in the case of a Gaussian modulation. 
For other modulation schemes, the operator $\hat{a}_\tau$ will be slightly different and therefore $|\alpha_k\rangle$ will in general no longer be an eigenstate. 
Let us write without loss of generality
\[ \hat{a}_\tau |\alpha_k\rangle = u_k |\alpha_k\rangle + v_k |\alpha_k^\perp\rangle,\]
where $|\alpha_k^\perp\rangle$ is orthogonal to $|\alpha_k\rangle$ and $u_k, v_k$ are complex numbers.
We get
\[
\langle \alpha_k | a_\tau^\dag a_\tau |\alpha_k\rangle - |\langle \alpha_k | a_\tau |\alpha_k\rangle|^2 = | u_k|^2 + |v_k|^2 - |u_k|^2 = |v_k|^2
\]
and therefore
\begin{align*}
w =  \sum_k p_k \| \Pi_k^\perp \hat{a}_\tau |\alpha_k\rangle\|^2
\end{align*}
where $\Pi_k^\perp = \1 - |\alpha_k\rangle\langle \alpha_k|$ is the projector onto the subspace orthogonal to $|\alpha_k\rangle$.
In other words, $w$ quantifies how much weight from a random input state is mapped by $\hat{a}_\tau$ to an orthogonal subspace.

%%%%%%%%%%%%%%%%%%%%%%%%%%%%%%%%%%%%%%%%
\section{The $M$-PSK modulation}
\label{sec:PSK}

The goal of this section is to provide an explicit expression for the value of $Z^*$ of Eqn.~\eqref{eqn:bosonic} corresponding to the case of a lossy and noisy Gaussian channel:
\[ Z^*(T,\xi) = 2\sqrt{T} \; \tr(\tau^{1/2} a \tau^{1/2} a^\dag) -  \sqrt{2T\xi w}.\]
The state $\tau$ takes the following form for an $M$-PSK modulation consisting of the states $|\alpha e^{ik\theta}\rangle$ for $\theta = 2\pi/M$ and $\alpha >0$:
\[ \tau = \frac{1}{M} \sum_{k=0}^{M-1} |\alpha e^{ik\theta}\rangle \langle \alpha e^{ik\theta}| = e^{-\alpha^2} \sum_{k=0}^{M-1} \nu_k |\phi_k\rangle \langle \phi_k|,\]
with 
\begin{align*}
|\phi_k\rangle = \frac{1}{\sqrt{ \nu_k}} \sum_{n=0}^\infty \frac{\alpha^{nM +k}}{\sqrt{(nM+k)!}} |nM+k\rangle,
\end{align*}
and
\begin{align*}
\nu_k = \sum_{n=0}^\infty \frac{\alpha^{2(nM +k)}}{(nM+k)!}.
\end{align*}
This expression for $\nu_k$ involves an unnecessary infinite sum and can be simplified. 
Let us introduce $\mu_j$ which is obtained by applying a discrete Fourier transform
\[ \mu_j := \sum_{k=0}^{M-1}  e^{i j k \theta} \nu_k  =\sum_{k=0}^{M-1} \sum_{n=0}^\infty  e^{i j k \theta} \frac{\alpha^{2(nM +k)}}{(nM+k)!}  = \sum_{m=0}^{\infty} e^{i j m \theta} \frac{\alpha^{2m}}{m!} =  \exp(\alpha^2 e^{ij\theta}),\]
where we used that $e^{i j n\theta} = e^{ij (n \, \mathrm{mod} M) \theta}$. Applying an inverse Fourier transform gives:
\[ \nu_k =\frac{1}{M} \sum_{j=0}^{M-1} e^{-ijk \theta}  \exp(\alpha^2 e^{ij\theta}).\]

We now wish to compute $\tr(\tau^{1/2} a \tau^{1/2} a^\dag)$. It is straightforward to check that:
\[ \langle \phi_j |\alpha_k\rangle = e^{-\alpha^2/2}\sqrt{\nu_j} e^{ijk\theta} \qquad \text{and} \qquad a |\phi_k\rangle = \alpha \frac{\nu_{k-1}^{1/2}}{\nu_k^{1/2}} |\phi_{k-1}\rangle,\]
where indices are taken modulo $M$.
This gives
\begin{align*}
\tr(\tau^{1/2} a \tau^{1/2} a^\dag) &= e^{-\alpha^2} \sum_{k,\ell=0}^{M-1} \sqrt{\nu_k \nu_\ell} \langle \phi_k | a |\phi_\ell\rangle \langle \phi_\ell | a^\dag |\phi_k\rangle\\
&= \alpha^2 e^{-\alpha^2} \sum_{k,\ell=0}^{M-1} \sqrt{\nu_k \nu_\ell} \frac{\nu_{\ell-1}}{\nu_\ell}|\langle \phi_k  |\phi_{\ell-1}\rangle |^2\\
&= \alpha^2 e^{-\alpha^2} \sum_{k=0}^{M-1}\frac{\nu_k^{3/2}}{\nu_{k+1}^{1/2}}
\end{align*}
where the last equality results from the orthogonality of the $\{|\phi_k\rangle\}$ family.
The operator $a_\tau = \tau^{1/2} a \tau^{-1/2}$ takes a simple form:
\begin{align*}
a_\tau = \sum_{k,\ell=0}^{M-1} \frac{\nu_k^{1/2}}{\nu_\ell^{1/2}} |\phi_k\rangle \langle \phi_k|a|\phi_\ell\rangle \langle \phi_\ell| =\alpha  \sum_{k=0}^{M-1} \frac{\nu_k}{\nu_{k+1}}  |\phi_k\rangle \langle \phi_{k+1}|.
\end{align*}
We can finally compute $w$:
\begin{align*}
w & =  \sum_k p_k \left( \langle \alpha_k | a_\tau^\dag a_\tau |\alpha_k\rangle - |\langle \alpha_k | a_\tau |\alpha_k\rangle|^2\right)\\
&=\frac{1}{M} \sum_{k=0}^{M-1} \langle \alpha_k | \alpha^2 \left(\sum_{j=0}^{M-1} \frac{\nu_j^2}{\nu_{j+1}^2} |\phi_{j+1}\rangle \langle \phi_{j+1}| \right)|\alpha_k\rangle- \frac{\alpha^2}{M}  \sum_{k=0}^{M-1}\left|  \sum_{j=0}^{M-1} \frac{\nu_j}{\nu_{j+1}} \langle\alpha_k  |\phi_j\rangle \langle \phi_{j+1}| \alpha_k\rangle\right|^2\\
&=\frac{\alpha^2}{M} \sum_{k=0}^{M-1}\sum_{j=0}^{M-1}  \frac{\nu_j^2}{\nu_{j+1}^2} \langle \alpha_k |\phi_{j+1}\rangle \langle \phi_{j+1}| \alpha_k\rangle- \frac{\alpha^2}{M} e^{-2\alpha^2} \sum_{k=0}^{M-1}\left( \sum_{j=0}^{M-1} \frac{\nu_j^{3/2}}{\nu_{j+1}^{1/2}}  \right)^2\\
&=\alpha^2 e^{-\alpha^2} \sum_{j=0}^{M-1} \frac{\nu_j^2}{\nu_{j+1}} - \alpha^2 e^{-2\alpha^2} \left(\sum_{j=0}^{M-1}  \frac{\nu_j^{3/2}}{\nu_{j+1}^{1/2}}  \right)^2.
\end{align*}
Putting these results together, we obtain the following value for $Z^*(T, \xi)$ for a general $M$-PSK modulation:
\begin{align}\label{eqn:Z-PSK}
Z^*(T,\xi) = \sqrt{T} \left(2 \alpha^2 e^{-\alpha^2} \sum_{k=0}^{M-1}\frac{\nu_k^{3/2}}{\nu_{k+1}^{1/2}} -  \sqrt{2\xi \alpha^2 } \sqrt{  e^{-\alpha^2} \sum_{j=0}^{M-1} \frac{\nu_j^2}{\nu_{j+1}} -  e^{-2\alpha^2} \left(\sum_{j=0}^{M-1}  \frac{\nu_j^{3/2}}{\nu_{j+1}^{1/2}}  \right)^2}\right).
\end{align}

\begin{figure}[!h]
\begin{center}
\includegraphics[trim=10 0 0 0,clip,scale=0.6]{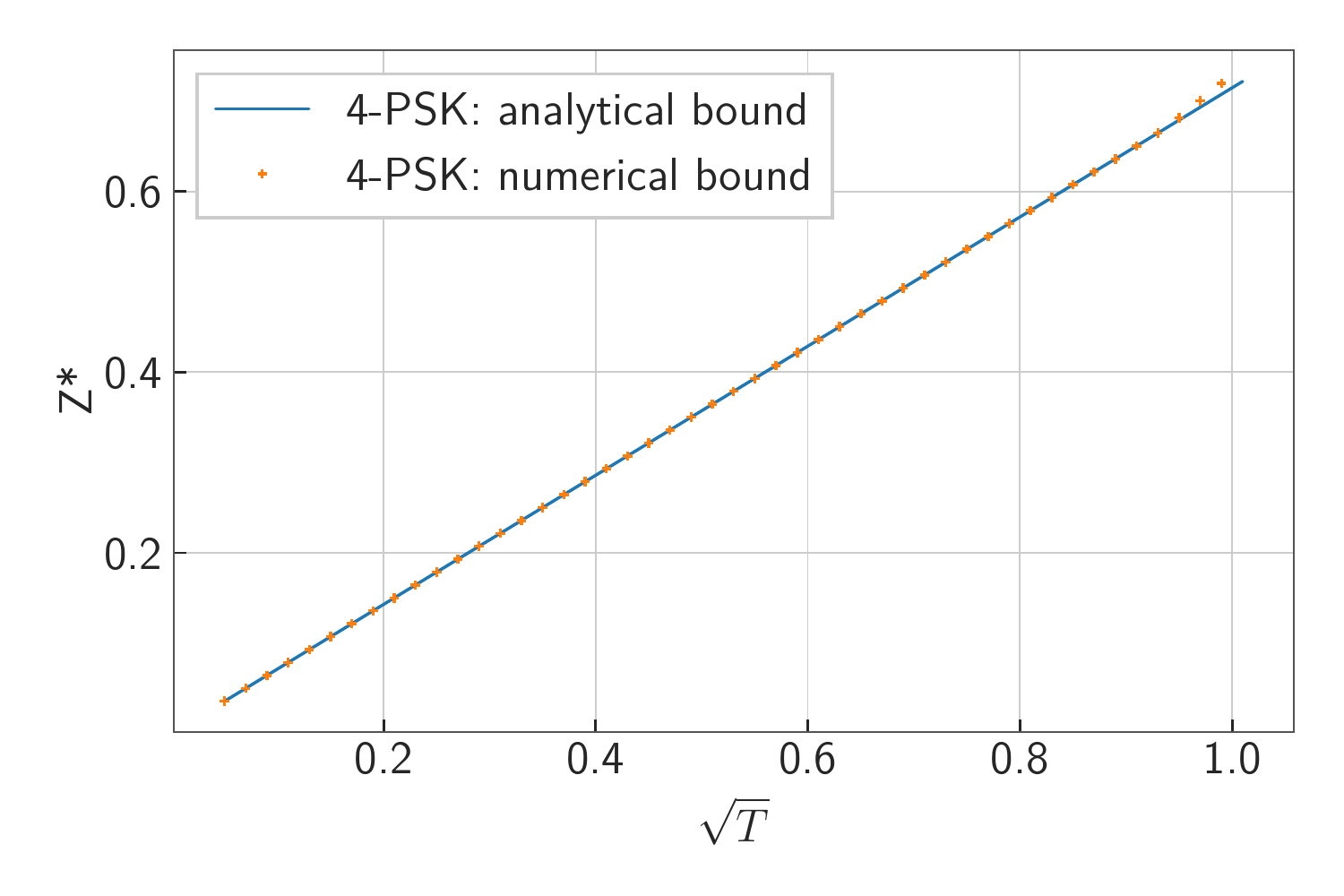}
\end{center}
\caption{Comparison between $Z^*(T,\xi)$ computed with Eqn.~\eqref{eqn:Z-PSK} for the $4$-PSK modulation, and the numerical result obtained by the SDP solver (as in Ref.~\cite{GGD19}), for $\alpha = 0.35$, $\xi=0.01$, as a function of the transmittance $T$. They match up to numerical precision, except for transmittances very close to 1, that are not relevant for experiments.}
\label{fig:Znum-vs-Ztheo}
\end{figure}

We compare in Fig.~\ref{fig:Znum-vs-Ztheo} our analytical bound with the numerical bound obtained in Ref.~\cite{GGD19}. We observe that they match up to numerical precision, except in the regime of very low-loss and large excess noise. While this regime is not very relevant for experiments, it would still be interesting to understand how to improve our numerical bound in that case. The question is whether there exists a better ansatz than that of Eqn.~\eqref{eqn:ansatz} more suited to this specific regime.

As we will see in Section \ref{sec:num}, the performance of the $M$-PSK protocols when using the above formula is essentially optimal for $M = 4$. In fact, the increase in performance when going to $M=5$ is very small and $M=6$ already reaches the asymptotic limit $M\to \infty$.
Of course, it is quite possible that this is only an artefact of our reliance on the extremality of Gaussian states and that the approach of \cite{LUL19} may show that larger values of $M$ are indeed useful.

%%%%%%%%%%%%%%%%%%%%%%%%%%%%%%%%%%%%%%%%
\section{General constellations}
\label{sec:constellation}

The conclusion of these previous sections is that the bound we obtain for the SDP is indeed tight in the two extreme cases where the constellation is either very small (as in $M$-PSK) or infinitely large (as in the Gaussian case). For constellations that fall in between, such as the general QAM that we will discuss now, it is not possible to compare our results to any numerical data (since none is available), but it is tempting to conjecture that our bound will likely be close to optimal.

The main lesson one can draw from the formula obtained in Eqn.~\eqref{eqn:bosonic} for $Z$ is that the key rate will increase when the modulation scheme gets closer to a Gaussian distribution, and this is mainly quantified by the value of 
\[ w=  \sum_k p_k \left( \langle \alpha_k | a_\tau^\dag a_\tau |\alpha_k\rangle - |\langle \alpha_k | a_\tau |\alpha_k\rangle|^2\right).
\]
There exist many choices of constellations that can be used to approximate a Gaussian distribution. For instance, the Gaussian quadrature rule is designed to match the first moments of the Gaussian distribution and works well for large constellations. The binomial (or random walk distribution) works much better for small constellations \cite{WV10,LRS16} and provides a natural candidate for CV QKD applications.

The normalized random walk distribution contains $m$ points for each quadrature, which are equally spaced between $-\sqrt{m-1}$ and $\sqrt{m-1}$, with associated probabilities corresponding to the binomial distribution. We choose a variance per coordinate equal to $\alpha^2/2$, which translates into $ \tr( \tau \hat{x}^2) = \tr(\tau \hat{p}^2) = 2\alpha^2 = V_A$ with our convention that $[\hat{x},\hat{p}]=2i$. The $M=m^2$ coherent states $|\alpha_{k,\ell}\rangle$ of the modulation are of the form
\begin{align}\label{eqn:bin}
\alpha_{k,\ell} =  \frac{\alpha\sqrt{2}}{\sqrt{m-1}} \left(k -\frac{m-1}{2}\right) + i  \frac{\alpha \sqrt{2}}{\sqrt{m-1}} \left(\ell -\frac{m-1}{2}\right),
\end{align}
chosen with probability
\begin{align}\label{eqn:bin2} 
p_{k,\ell} =  \frac{1}{2^{2(m-1)}} \tbinom{m-1}{k} \tbinom{m-1}{\ell}.
\end{align}

Another simple distribution is the discrete Gaussian distribution, where the coherent states are centered at $m^2$ possible equidistant points of the form $\alpha = x + i p$, with a respective probability given by
\begin{align} \label{eqn:discrete-gauss}
p_{x,p} \sim \exp\Big(- \nu (x^2+p^2)\Big).
\end{align}
This distribution is characterized by $\nu>0$ and by the spacing between the possible values of $x$ (or $p$). This spacing is, however, constrained once we fix the overall variance to $\alpha^2/2$ per coordinate. We are then left with a single parameter $\nu$ that can be optimized to maximize the secret key rate. 

As we will discuss in more detail in Section \ref{sec:num}, the two modulation schemes yield very close performance for QAM of size 64 or above, once the parameters of the discrete Gaussian distribution have been optimized. For simplicity, it is therefore more convenient to use the binomial distribution which comes without extra-optimization step. However, for smaller constellations, like 16-QAM, it seems that the discrete Gaussian distribution gives better results, and it would be interesting to find out whether other distributions are even better.

%%%%%%%%%%%%%%%%%%%%%%%%%%%%%%%%%%%%%%%%

\section{Modulation of arbitrary states}
\label{sec:thermal}

Our approach extends to the case where Alice sends arbitrary states $\tau_k$, with probability $p_k$, for instance squeezed states \cite{CLV01} or thermal states \cite{fil08,UF10}. Besides possible applications such as the application of CV QKD to the microwave regime \cite{WPL10}, it is important to be able to analyse the security of the protocol when the state preparation is imperfect since Alice can never prepare the intended states with infinite precision.
As an example, a modulation of thermal states consists in sending some displaced thermal state $\tau_k$ with $\langle n \rangle_{\mathrm{th}}$ photons centered around $\alpha_k$ with probability $p_k$. The state $\tau_k$ is given by
\[ \tau_k = D_{\alpha_k} \rho_{\mathrm{th}} D_{\alpha_k}^\dag \quad \text{with} \quad \rho_{\mathrm{th}} = \frac{1}{1+\langle n \rangle_{\mathrm{th}}}  \sum_{m=0}^\infty \left(\frac{\langle n \rangle_{\mathrm{th}}}{1+\langle n \rangle_{\mathrm{th}}}\right)^m |m\rangle \langle m|,\]
where $\rho_{\mathrm{th}}$ is a thermal state centered in phase space and $D_{\alpha_k} := \exp(\alpha_k \hat{b}^\dag - \bar{\alpha_k} \hat{b})$ is the operator describing a displacement by $\alpha_k$.

In this section, we will therefore consider the most general setting where Alice picks some index $k$ with probability $p_k$ and sends some state $\tau_k$, which is arbitrary. 
The security analysis relies on the same idea as before, that is computing the covariance matrix of the state $\rho_{AB}$ shared by Alice and Bob in the entanglement-based (EB) version of the protocol, and the covariance term can again be bounded with an SDP similar to Eqn.~\eqref{sdp:primal}.

The modulation is still characterized by its average state 
\begin{align}\label{eqn:tau-gen}
\tau := \sum_k p_k \tau_k
\end{align}
and we will keep the same purification as before to analyze the EB version of the protocol: 
\[ |\Phi\rangle_{AA'} := (\1 \otimes \tau^{1/2}) \sum_{n=0}^\infty |n\rangle_A |n\rangle_{A'}.\]
We need to replace the rank-one projector $|\psi_k\rangle \langle \psi_k|$ defined in Eqn.~\eqref{eqn:psi} by a positive semidefinite operator 
\begin{align}\label{eqn:Pk}
P_k := p_k \bar{\tau}^{-1/2} \bar{\tau_k} \bar{\tau}^{-1/2}.
\end{align}
These operators yield a resolution of the identity on the support of $\bar{\tau}$, the complex conjugate of $\tau$ (also equal to the transpose $\tau^{T}$ with respect to the Fock basis):
\[ \sum_k P_k = \Pi,\]
where $\Pi$ is the projector onto the support of $\bar{\tau}$. Since $\bar{\tau} = \tr_{A'} (|\Phi\rangle\langle \Phi|)$ corresponds to the reduced state on the system $A$, we can interpret the family $\{P_k\}$ as the POVM elements of a general measurement performed by Alice on $A$: whenever she obtains the measurement outcome $k$, the state of system $A'$ collapses to $\tau_k$.

Recall that the first-moment values that can be measured in the PM protocol are 
\[ c_1 = \mathrm{Re}( {\bm \alpha_\tau} | {\bm\beta}), \qquad c_2 = \mathrm{Re}({\bm\alpha} | {\bm \beta}),\]
with ${\bm \alpha_\tau} =  (\tr (\tau_k a_\tau))_k$.
These can be expressed as the expectation values of $\rho$ for the observables $C_1$ and $C_2$ defined by
\begin{align}\label{eqn:Cth}
C_1 &:= \sum_k z_k P_k \otimes b +\text{h.c.}\\
C_2 & := \sum_k \bar{\alpha_k} P_k \otimes b + \mathrm{h.c.}
\end{align}
with
\begin{align}\label{eqn:zk}
z_k := \tr(\bar{\tau_k} a_{\bar{\tau}}).
\end{align}
We also introduce the operators $G_1, G_2$ acting on the system $A$:
\begin{align}\label{eqn:Gi}
G_1 &:= \sum_k  z_k P_k, \quad G_2  := \sum_k \bar{\alpha_k} P_k 
\end{align}
and observe that 
\[ C_1 = G_1 \otimes b + \mathrm{h.c} \quad \text{and} \quad C_2 = G_2 \otimes b + \mathrm{h.c.}\]
We can now give the relevant SDP when we consider a modulation of arbitrary states:
\begin{align}
\min \quad & \tr(\rho\, C) \label{sdp:thermal}\\
\mathrm{s.t.} \quad  &\left\{   \begin{array}{l}\tr_B(\rho) = \bar{\tau} \nonumber\\
\tr\Big(\rho \, C_1 \Big) =2 c_1 \nonumber\\
\tr\Big(\rho \,  C_2 \Big) = 2 c_2, \nonumber\\
\tr( \rho (\Pi \otimes \hat{b}^\dag \hat{b})) = n_B,\nonumber\\
\rho \succeq 0.
\end{array}\right.
\end{align}

Our goal is again to exhibit operators $K_{\pm}$ and exploit the operator inequalities $K_{\pm} K_{\pm}^\dag \succeq 0$ to bound the value of the SDP. We need some additional notations:
\begin{align}
A &:= \sum_k \langle k| \otimes \Pi a P_k^{1/2} \otimes D_{t \alpha_k}\\
B &:= \sum_{k,\ell} |k\rangle \otimes P_k^{1/2} P_\ell \otimes D_{t\alpha_k}^\dag (b-t\alpha_\ell)\\
F &:= \sum_{k} z_k \langle k| \otimes P_k^{1/2} \otimes D_{t\alpha_k}
\end{align}
where $\{|k\rangle\}$ is an orthonormal basis of a reference system $R$, storing Alice's measurement result. 
The operators $A$ and $B$ should not be confused with the registers $A$ and $B$. We recall that the operator $D_{\beta}$ describes a displacement by $\beta$.

We then proceed exactly as in Section \ref{sec:primal} and define 
\[ K_{\pm} := z(A - F) \pm \frac{1}{z} B^\dag.\]
Considering $K_\pm K_\pm^\dag \succeq 0$ results in the sum-of-squares inequality:
\begin{align}\label{eqn:sos9}
\underbrace{z^2 (A-F)(A-F)^\dag}_{(1)}  \pm \underbrace{(AB + B^\dag A^\dag)}_{(2)} \mp \underbrace{(FB +B^\dag F^\dag)}_{(3)} + \underbrace{\frac{1}{z^2} B^\dag B}_{(4)} \succeq 0.
\end{align}
We take the expectation with respect to the state $\rho$ and consider each term individually. 

\begin{enumerate}
\item For the first term, we have
\begin{align*}
AA^\dag &= \Pi a \Pi a^\dag \Pi\\
AF^\dag &= \sum_k \bar{z}_k \Pi a P_k\\
FF^\dag&= \sum_k |z_k|^2 P_k.
\end{align*}
Their expectation with respect to $\rho$ gives
\begin{align*}
\tr(\rho AA^\dag) &= \tr(\bar{\tau} a \Pi a^\dag) = \tr(\tau a_\tau^\dag a_\tau)\\
\tr(\rho AF^\dag) &= \sum_k \bar{z}_k \tr(\bar{\tau}  a P_k) = \sum_k p_k |z_k|^2\\
\tr(\rho FF^\dag)&= \sum_k |z_k|^2 \tr(\bar{\tau} P_k) = \sum_k p_k |z_k|^2.
\end{align*}
Putting everything together, we get
\begin{align}\label{eqn:term1th}
\tr( \rho \cdot (1) ) = z^2 w, 
\end{align}
where we define 
\[ w := \tr(\tau a_\tau^\dag a_\tau) -\sum_k p_k |\tr(\tau_k a_\tau)|^2.\]

\item For the second term, we have
\[ AB = \sum_k \Pi a P_k \otimes (b-t\alpha_k) \]
and the expectation with respect to $\rho$ gives
\[ \tr(\rho AB) = \tr(\rho(\Pi a \Pi \otimes b) ) - t \sum_k \alpha_k \tr(\bar{\tau} a P_k) = \tr(\rho(\Pi a \Pi \otimes b) ) - t \sum_k p_k \alpha_k z_k .\]
In particular, we can recognize the objective function of the SDP:
\begin{align}\label{eqn:term2th}
\tr( \rho \cdot (2) ) = \tr(\rho \, C) - 2 t \mathrm{Re} \Big(\sum_k p_k \alpha_k z_k\Big).
\end{align}

\item For the third term of Eqn.~\eqref{eqn:sos9}, we note that
\begin{align*}
F B &= \sum_{k, \ell}  z_k P_k P_\ell \otimes (b - t\alpha_\ell)\\
&= G_1 \otimes b - t \sum_{k,\ell} z_k \alpha_\ell P_k P_\ell\\
&= G_1 \otimes b - t G_1 G_2^\dag.
\end{align*}
The expectation with respect to $\rho$ gives
\begin{align}\label{eqn:term3th}
\tr(\rho \cdot (3) ) = 2 c_1 - 2t \mathrm{Re} \Big(\sum_{k,\ell} z_k \alpha_\ell \tr(\bar{\tau} P_k P_\ell)\Big).
\end{align}

\item Finally, for the fourth term, we have
\begin{align*}
B^\dag B &= \sum_{k,\ell} P_k P_\ell \otimes (b-t\alpha_k)^\dag (b-t\alpha_\ell)\\
&= \Pi \otimes b^\dag b - t \sum_k P_k \otimes ( \bar{\alpha_k} b + \alpha_k b^\dag) + t^2 \sum_{k, \ell} \bar{\alpha_k} \alpha_\ell P_k P_\ell\\
&=\Pi\otimes  b^\dag b - t (G_2 \otimes b + G_2^\dag \otimes b^\dag) + t^2 G_2^\dag G_2.
\end{align*}
The expectation with respect to $\rho$ gives
\begin{align}\label{eqn:term4th}
\tr(\rho \cdot (4) ) = \frac{1}{z^2} \Big( n_B - 2 tc_2  + t^2 \tr(\bar{\tau} G_2^\dag G_2)\Big).
\end{align}

\end{enumerate}

By considering the four terms of Eqn.~\eqref{eqn:sos9}, we find that
\begin{align*}
0 &\leq z^2 w \pm \Big(\tr(\rho\,  C) - 2 t \mathrm{Re} \Big(\sum_k p_k \alpha_k z_k\Big) \Big) \mp 2\Big( c_1 - t \mathrm{Re} \Big(\sum_{k,\ell} z_k \alpha_\ell \tr(\bar{\tau} P_k P_\ell)\Big) \Big) \\
& \quad + \frac{1}{z^2} \Big(n_B - 2 tc_2  + t^2 \tr(\bar{\tau} G_2^\dag G_2) \Big)\\
&=  z^2 w \pm \Big(\tr(\rho\, C) - 2 t \mathrm{Re} \Big(\sum_{k,\ell} (\alpha_k - \alpha_\ell) z_k  \tr(\bar{\tau} P_k P_\ell)\Big) -2c_1 \Big) + \frac{1}{z^2} \Big(n_B - 2 tc_2  + t^2 \tr(\bar{\tau} G_2^\dag G_2) \Big)
\end{align*}
where we used the substitution $p_k = \sum_\ell \tr(\bar{\tau} P_k P_\ell)$ in the second equality.
Overall, this implies the two inequalities
\begin{align*}
\tr(\rho\, C) &\leq 2 c_1 + 2t \, \mathrm{Re}\Big(\sum_{k,\ell} (\alpha_k - \alpha_\ell) z_{k}  \tr(\bar{\tau} P_k P_\ell) \Big) + 2 \sqrt{w  \Big(n_B + t^2  \tr(\bar{\tau} G_2^\dag G_2) -  2 tc_2\Big)}\\
\tr(\rho\, C) &\geq 2 c_1 + 2t \,\mathrm{Re}\Big(\sum_{k,\ell} (\alpha_k - \alpha_\ell) z_{k}  \tr(\bar{\tau} P_k P_\ell)\Big) - 2 \sqrt{w  \Big(n_B + t^2  \tr(\bar{\tau} G_2^\dag G_2) -  2 tc_2\Big)}.
\end{align*}
where we optimized the variable $z$ exactly as in Section \ref{sec:primal}.
We note a potential problem in the case where $w$ vanishes: it would then appear that by fixing $t$ arbitrarily, we could obtain any bound about on $\tr(\rho \, C)$. This is not possible, however, since $w$ only vanishes for a Gaussian modulation of coherent states and in this case the second term of the right-hand side also vanishes. More generally, this term vanishes whenever the measurement performed by Alice is projective, in the sense that $P_k P_\ell = \delta_{k,\ell} P_k$, corresponding for instance to an arbitrary modulation of coherent states (or pure squeezed states).
Here, we simply choose the value of $t$ that minimizes the term under the square-root (but note that this may be suboptimal in general), namely
\[ t = \frac{c_2}{ \tr(\bar{\tau} G_2^\dag G_2) } = \frac{c_2}{\sum_{k,\ell} \bar{\alpha_k} {\alpha_\ell}  \tr(\bar{\tau} P_k P_\ell)}.\]
This establishes our final bounds:
\begin{align*}\label{eqn:final-upp}
\tr(\rho\, C) &\geq 2 c_1 - 2c_2  \frac{  \mathrm{Re}\Big(\sum_{k,\ell} (\alpha_\ell - \alpha_k) z_{k}  \tr(\bar{\tau} P_k P_\ell) \Big) }{ \tr(\bar{\tau} G_2^\dag G_2)}- 2 \sqrt{w  \left(n_B - \frac{ c_2^2}{  \tr(\bar{\tau} G_2^\dag G_2)}\right)},\\
\tr(\rho\, C) &\leq 2 c_1 - 2c_2  \frac{  \mathrm{Re}\Big(\sum_{k,\ell} (\alpha_\ell - \alpha_k) z_{k}  \tr(\bar{\tau} P_k P_\ell) \Big) }{ \tr(\bar{\tau} G_2^\dag G_2)}+ 2 \sqrt{w  \left(n_B - \frac{ c_2^2}{  \tr(\bar{\tau} G_2^\dag G_2)}\right)}.
\end{align*}
It would be interesting to understand whether this lower bound is tight for a Gaussian modulation of thermal states. 

%%%%%%%%%%%%%%%%%%%%%%%%%%%%%%%%%%%%%%%%
\section{Finite-size effects}
\label{sec:finite}

In this section, we quickly discuss two of the main finite-size effects that will need to be included in a future full composable security proof against general attacks. Another important effect concerns the optimality of collective attacks among general attacks. At the moment, this point still needs to be clarified, and we leave it for future work. Note, however, that the correction term due to this last effect is typically dependent on the proof techniques and we have observed in the past that better techniques can significantly reduce this term. For instance for DV QKD, the first techniques were based on the exponential de Finetti theorem \cite{ren07}, then on a de Finetti reduction \cite{CKR09}, then on an entropic uncertainty principle \cite{TR11} and finally on the entropy accumulation theorem \cite{DFR20}. It is therefore tempting to believe that a similar phenomenon will occur with CV QKD, and this has indeed been the case for protocols with a Gaussian modulation of coherent states where both an exponential de Finetti theorem \cite{RC09} and a Gaussian de Finetti reduction \cite{lev17} are known. 

For these reasons, it makes sense to focus on the two finite-size effects that will likely remain the dominating terms in any future full security proof of CV QKD, namely parameter estimation and reconciliation efficiency.

\subsection{Parameter estimation}
\label{sec:PE}

One of the novelties of our proof, when compared to the case of a Gaussian modulation, is the need for experimentally estimating 3 parameters, $c_1$, $c_2$ and $n_B$, in order to get an upper bound on the Holevo information $\chi(Y;E)_\rho$ appearing in the Devetak-Winter bound. Let us denote by $f(c_1, c_2, n_B)$ this upper bound, which is given explicitly in Eqn.~\eqref{eqn:chiYE}, where we compute the symplectic eigenvalues for the covariance matrix $\Gamma' = \left[ \begin{smallmatrix} V\1_2 & Z^* \sigma_Z\\Z^* \sigma_Z & W \1_2\end{smallmatrix} \right]$ with $V$ given by the modulation scheme, $W$ computed from the value of $n_B$ and $Z^*$ computed from the values of $c_1, c_2, n_B$ by the formula given in Eqn.\eqref{eqn:Z^*}.
We note that the function $f$ depends implicitly on the modulation scheme, for example \textit{via} the value of $w$ appearing in the expression of $Z^*$.

Since $n_B$ is the average photon number in Bob's system, it corresponds to the variance (up to a shift and a factor 2) of his quadrature measurements, when the distribution is centered:
\[ 1+2n_B =  1+2 \tr(\rho b^\dag b) = \frac{1}{2} \Big(\langle \hat{x}_B^2 \rangle_\rho + \langle \hat{p}_B^2\rangle_\rho \Big).\]
One can then compute an observed value $n_B^{\text{obs}}$ corresponding to the empirical average of $n_B$ evaluated on the samples that are used for parameter estimation.
In order to estimate $c_1$ and $c_2$, one can for instance form a vector of average observed values ${\bm \beta}^{\text{obs}} = (\beta_k^{\text{obs}})_k$ where $\beta_k^{\text{obs}}$ is the average observed outcome for the observable $\hat{b} = \frac{1}{{2}}(\hat{x}_B + i\hat{p}_B)$ when Alice has sent the state $|\alpha_k\rangle$, and then compute 
\[ c_1^{\text{obs}} := \mathrm{Re}( {\bm \alpha_\tau} | {\bm\beta}^{\text{obs}}), \qquad c_2^{\text{obs}} := \mathrm{Re}({\bm\alpha} | {\bm \beta}^{\text{obs}}),\]
where the $k^{\mathrm{th}}$ entry of the vectors ${\bm \alpha_\tau}$ and ${\bm \alpha}$ are given respectively by $ \langle \alpha_k |a_\tau |\alpha_k\rangle$ and $\alpha_k$.

In the asymptotic setting, one can assume that the values of $c_1$, $c_2$ and $n_B$ are known exactly, and therefore coincide with their observed values. This is not the case in the finite-size setting, and one would in general compute a confidence region for the triple $(c_1, c_2, n_B)$ compatible with the observed values $(c_1^{\text{obs}}, c_2^{\text{obs}}, n_B^{\text{obs}})$. One can check numerically that the function $f(c_1, c_2, n_B)$ is increasing with $n_B$ and decreasing with either $c_1$ or $c_2$, when the other 2 variables are fixed. This implies that there is no need for computing the whole confidence region, but it is in fact sufficient to compute ``worst-case estimates'' for $c_1, c_2$ and $n_B$, in the sense that 
\[ \mathrm{Pr} [c_1 \leq c_1^{\mathrm{min}}]\leq \frac{\eps_{\text{PE}}}{3}, \quad  \mathrm{Pr} [c_2 \leq c_2^{\mathrm{min}}]\leq \frac{\eps_{\text{PE}}}{3}, \quad  \mathrm{Pr} [n_B \geq n_B^{\mathrm{max}}]\leq \frac{\eps_{\text{PE}}}{3}.\]
In these expressions, the variables $c_1, c_2$ and $n_B$ refer to their respective values for the modes that have not been used for parameter estimation, and that will be exploited for key extraction. The numbers $c_1^{\mathrm{min}}, c_2^{\mathrm{min}}, n_B^{\mathrm{max}}$ are computed with Eqn.~\eqref{eqn:finite} below from observations made during the parameter estimation procedure and correspond to the worst-case estimators. The small parameter $\eps_{\text{PE}}$ is an upper bound on the probability that the parameter estimation performed by Alice and Bob returns $c_1^{\mathrm{min}}$ for instance and that the value of $c_1$ is less than $c_1^{\mathrm{min}}$ for the remaining unobserved modes. Once these numbers are known, one can simply use the following upper bound on $\chi(Y;E)$ in the Devetak-Winter bound:
\[ \chi(Y;E) \leq f(c_1^{\mathrm{min}}, c_2^{\mathrm{min}}, n_B^{\mathrm{max}}),\]
which holds, except with a small probability $\eps_{\text{PE}}$.

It is well known that such a parameter estimation is more subtle in the case of CV QKD because the random variables we aim at estimating are not trivially bounded by construction (contrary to the quantum bit error rate of BB84 for instance, which lies by definition between 0 and 1). This difficulty can be addressed with the tools developed in Ref.~\cite{lev15}, but this is beyond the scope of the present manuscript. 
Here, we simply wish to give the expected asymptotic scaling of $c_1^{\mathrm{min}}, c_2^{\mathrm{min}}$ and $n_B^{\mathrm{max}}$, as a function of $n$, the number of quantum states exchanged on the quantum channel:
\begin{align}\label{eqn:finite}
 n_B^{\mathrm{max}} = n_B^{\mathrm{obs}} \left(1 + O\left(\sqrt{\frac{\log(1/\eps_{\mathrm{PE}})}{n}}\right)\right), \qquad c_i^{\mathrm{min}} = c_i^{\mathrm{obs}} - O\left(n_B^{\mathrm{obs}}\sqrt{\frac{\log(1/\eps_{\mathrm{PE}})}{n}}\right), 
\end{align}
for $i\in \{1,2\}$.
The precise value of the hidden positive constants in the $O( \cdot)$ notation are not known at the moment, and will require a thorough analysis to determine.

\subsection{Reconciliation efficiency}
\label{sec:beta}

The information reconciliation step of the protocol is also more involved for CV QKD than for DV QKD. Without this step, or assuming it is achieved perfectly, the asymptotic secret key rate would read 
\begin{align}\label{eqn:beta=1}
 K = I(X;Y) - \sup_{\mathcal{N}} \chi(Y;E) = \inf_{\mathcal{N}} H(Y|E) - H(Y|X),
\end{align}
where $X$ and $Y$ denote the variables corresponding to Alice and Bob, and the raw key is given by Bob's variable (which is always the more favorable choice for CV QKD).
Since the present paper focusses on the asymptotic regime, one could in principle ignore the reconciliation procedure, but this would lead to incorrect predictions in the case of CV QKD because an imperfect reconciliation significantly affects the performance: for instance, with perfect reconciliation and a Gaussian modulation, the secret key rate is strictly increasing with the variance of the modulation, while this is no longer the case as soon as the reconciliation is slightly imperfect.

In a typical DV protocol, Alice and Bob hold correlated bit-strings $\vec{x} = (x_1, \ldots, x_n)$ and $\vec{y} = (y_1, \ldots, y_n)$ corresponding respectively to the input and output of $n$ uses of a binary symmetric channel, with crossing probability $p$. Bob then sends some side-information to Alice via the authenticated classical channel to help him recover the value of $\vec{y}$.
In the asymptotic limit where $n$ tends to infinity, the channel coding theorem ensures that Alice and Bob can succeed at this task with high probability provided that Alice sends $H(Y|X) = H(X|Y)= n h(p)$ bits of side information, with the binary entropy defined as $h(p):= -p \log_2 (p) - (1-p) \log_2 (1-p)$. 
In practice, one cannot achieve this perfectly, and Alice will need to send slightly more information, namely $(1 + f(p)) n h(p)$ bits, where $f(p)$ is typically a few percent.

For a CV QKD protocol, the relevant channel in practice\footnote{By relevant channel, we mean the channel that is typically observed in experimental implementations, and that therefore corresponds to a transmission in an optical fiber.} is the additive Gaussian white-noise (AWGN) channel: the strings held by Alice and Bob are $(x_1, \ldots, x_{n}) \in \mathbbm{C}^{n}$ and $(y_1, \ldots, y_{n})\in \mathbbm{C}^{n}$ where $x_i$ is chosen according to the modulation scheme: it is equal to $\alpha_k$ with probability $p_k$. For each $i$, we expect
\[ y_i = \sqrt{\frac{T}{2}} x_i + z_i,\]
where $\mathrm{Re}(z_i), \mathrm{Im}(z_i) \sim \mathcal{N}(0, 1 + T\xi)$ is a Gaussian noise. The extra factor $1/2$ in the square-root comes from the heterodyne detection which requires first splitting the incoming signal on a balanced beamsplitter before measuring each output mode with a homodyne detection.
In the case of a Gaussian modulation, with $\mathrm{Re}(x_i), \mathrm{Im}(x_i) \sim \mathcal{N}(0, V_A)$ two Gaussian random variables of variance $V_A$, the mutual information between the random variables $X$ and $Y$ takes a simple expression
\[ I(X;Y) = \log_2(1 + \mathrm{SNR}) \quad \text{with} \quad \mathrm{SNR} := \frac{TV_A}{2 + T\xi}.\]
Note that this is twice the standard formula $\frac{1}{2}\log_2(1 + \mathrm{SNR})$ because we consider both the real and imaginary parts.

For the modulation schemes we consider in this paper, there is no closed-form expression for the mutual information $I(X;Y)$, although it is typically very close to the Gaussian version, provided the variance $V_A$ is small enough \cite{WV10}. Note in particular, that for a $2^k$-QAM, it is necessarily upper bounded by $k$, which is itself an upper bound on the entropy $H(X)$, while $\log_2(1 + \mathrm{SNR})$ grows to infinity with the signal-to-noise ratio.
Assuming therefore that the gap between the two quantities is indeed negligible here, we still need to quantify how far we are from the key rate of Eqn.~\eqref{eqn:beta=1}.
There are two natural ways to write a version of the key rate taking into account the imperfect reconciliation efficiency:
\begin{align}\label{eqn:beta<1}
 K = \beta I(X;Y) - \sup_{\mathcal{N}} \chi(Y;E) = \inf_{\mathcal{N}} H(Y'|E) - (1+f) H(Y'|X),
\end{align}
where $\beta < 1$ is the so-called reconciliation efficiency generally used in CV QKD and $f > 0$ is more relevant to DV QKD. In the second expression, we write $Y'$ to denote a discretized version of $Y$, since otherwise the conditional entropy is ill-defined. 

Provided that the reconciliation protocol fully exploits soft-information, meaning that the discretization is sufficiently precise, then high values of $\beta$ between $95$ and $98\%$ are achievable \cite{JKL11,MCZ18,MGP21} for a Gaussian modulation. Similarly, for a QPSK modulation, it is possible to easily reach $90\%$ at arbitrarily low SNR. It is not clear, however, how to achieve similar numbers with a coarse graining corresponding to Bob simply keeping the sign of his variable in the QPSK case, as done in Ref.~\cite{LUL19}.

The reconciliation problem has not yet been studied in detail in the case of larger QAMs. Nevertheless, one can realistically assume that values around $95\%$ can be achieved, given the closeness between this problem and the Gaussian case. For this reason, we will assume  $\beta=0.95$ in the numerical simulations of Section \ref{sec:num}.

%%%%%%%%%%%%%%%%%%%%%%%%%%%%%%%%%%%%%%%%
\section{Numerical results}
\label{sec:num}

In this section, we perform some numerical simulations in the case of a typical Gaussian channel with transmittance $T$ and excess noise $\xi$. The covariance matrix $\Gamma'$ takes the form
\begin{align*}
\Gamma' := \begin{bmatrix}
(V_A+1) \1_2 & Z^* \sigma_Z\\
Z^* \sigma_Z & (1+ TV_A + T\xi) \1_2
\end{bmatrix}
\end{align*}
with 
\[ Z^* = 2\sqrt{T} \; \tr(\tau^{1/2} a \tau^{1/2} a^\dag) -  \sqrt{2T\xi w}\]
and $\tau$ and $W$ depend on the specific modulation scheme that is considered.

\begin{figure}[!h]
\begin{center}
\includegraphics[trim=0 0 0 2,clip,scale=0.485]{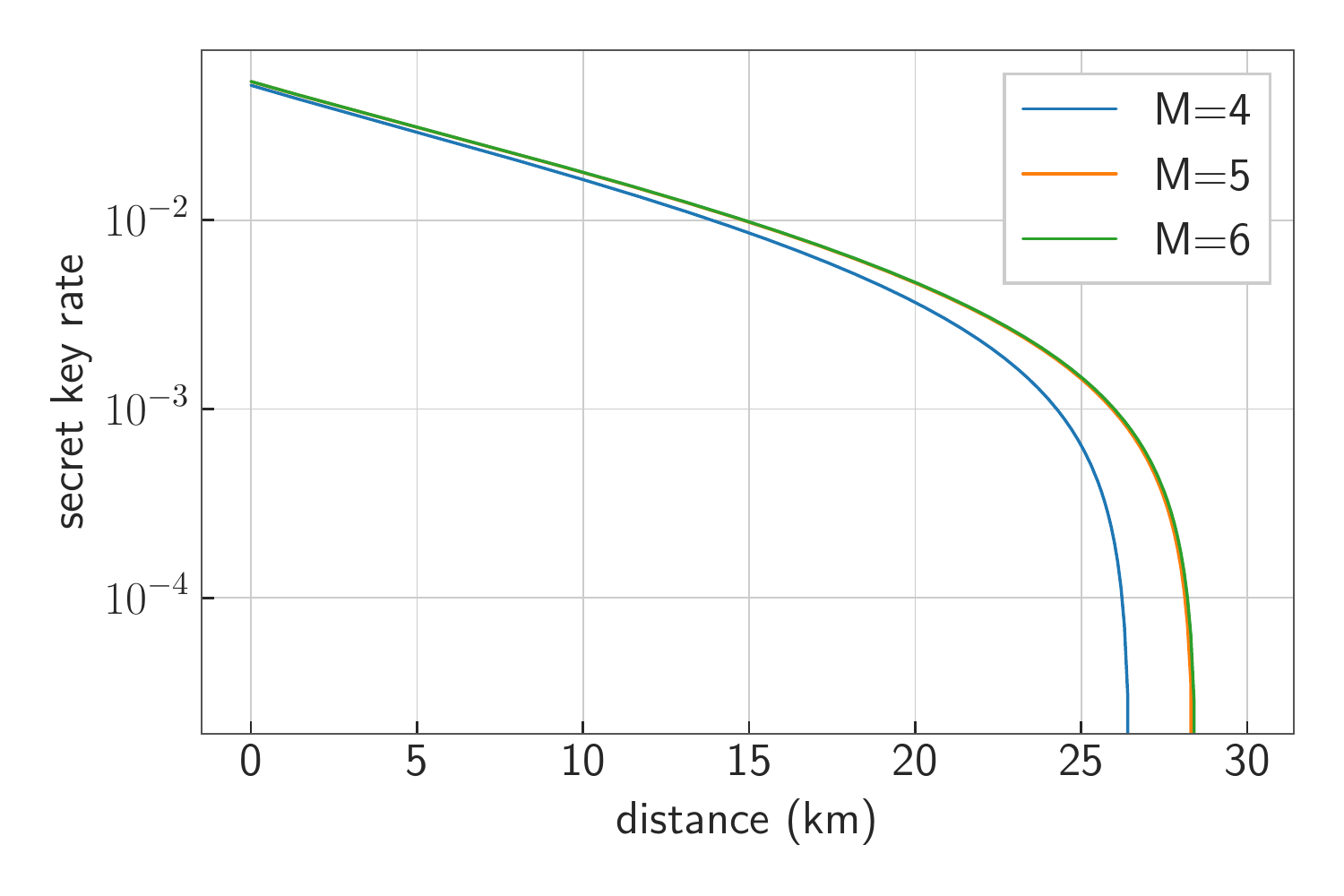} \includegraphics[trim=0 0 0 2,clip,scale=0.485]{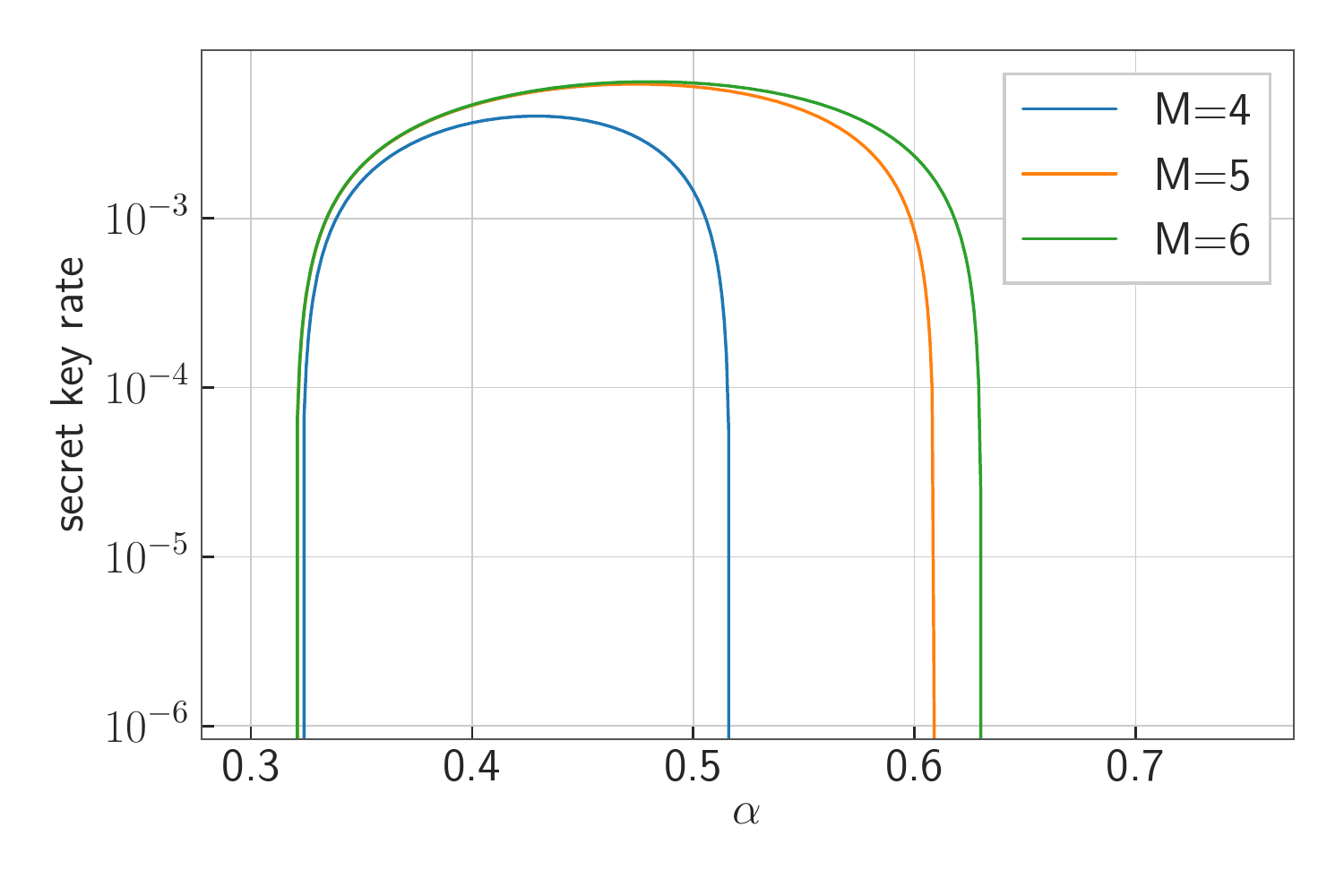}
\end{center}
\caption{Asymptotic secret key rate for the $M$-PSK modulation schemes with $M \in \{4,5,6\}$, from bottom to top. The other parameters are $\xi = 0.01$ and $\beta=0.95$. Left panel: the modulation variance is fixed, $\alpha = 0.4$, the rates for $M=5$ and $M=6$ are indistinguishable; right panel: secret key rate as a function of $\alpha$ for $d = 20$ km.}
\label{fig:456PSK}
\end{figure}

We first compare in Figure \ref{fig:456PSK} the secret key rates obtained for various sizes of the $M$-PSK modulation. 
The left panel shows that when the modulation variance (or equivalently, $\alpha$) is optimized, then going beyond $M=5$ is essentially useless. On the right panel, we see that the only advantage of increasing $M$ is to allow for larger possible values of $\alpha$. However, it is much better to consider QAM instead of increasing the number of states in the PSK modulation.

\begin{figure}[!h]
\begin{center}
\includegraphics[trim=0 15 0 2,clip,scale=0.485]{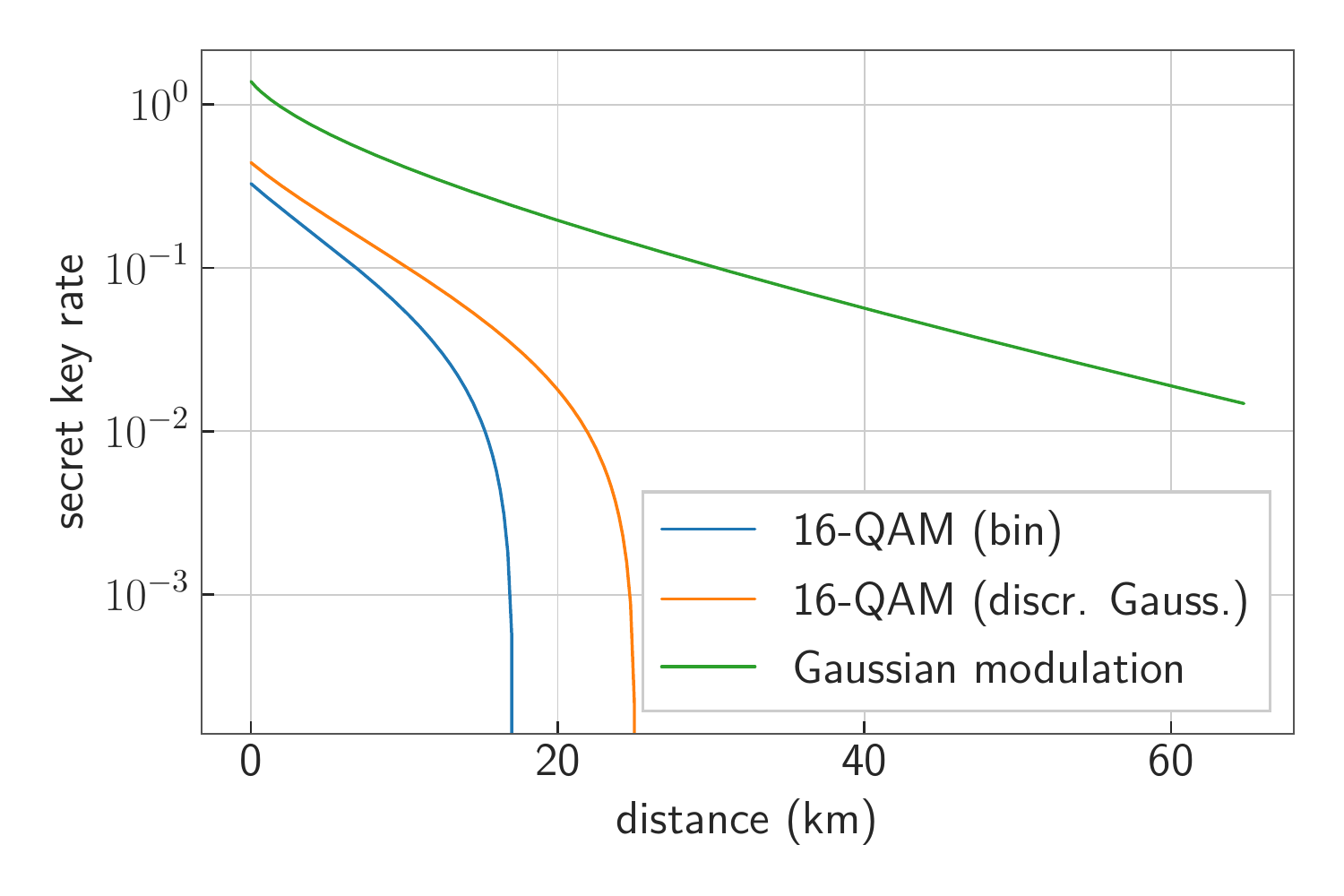}
\includegraphics[trim=0 15 0 2,clip,scale=0.485]{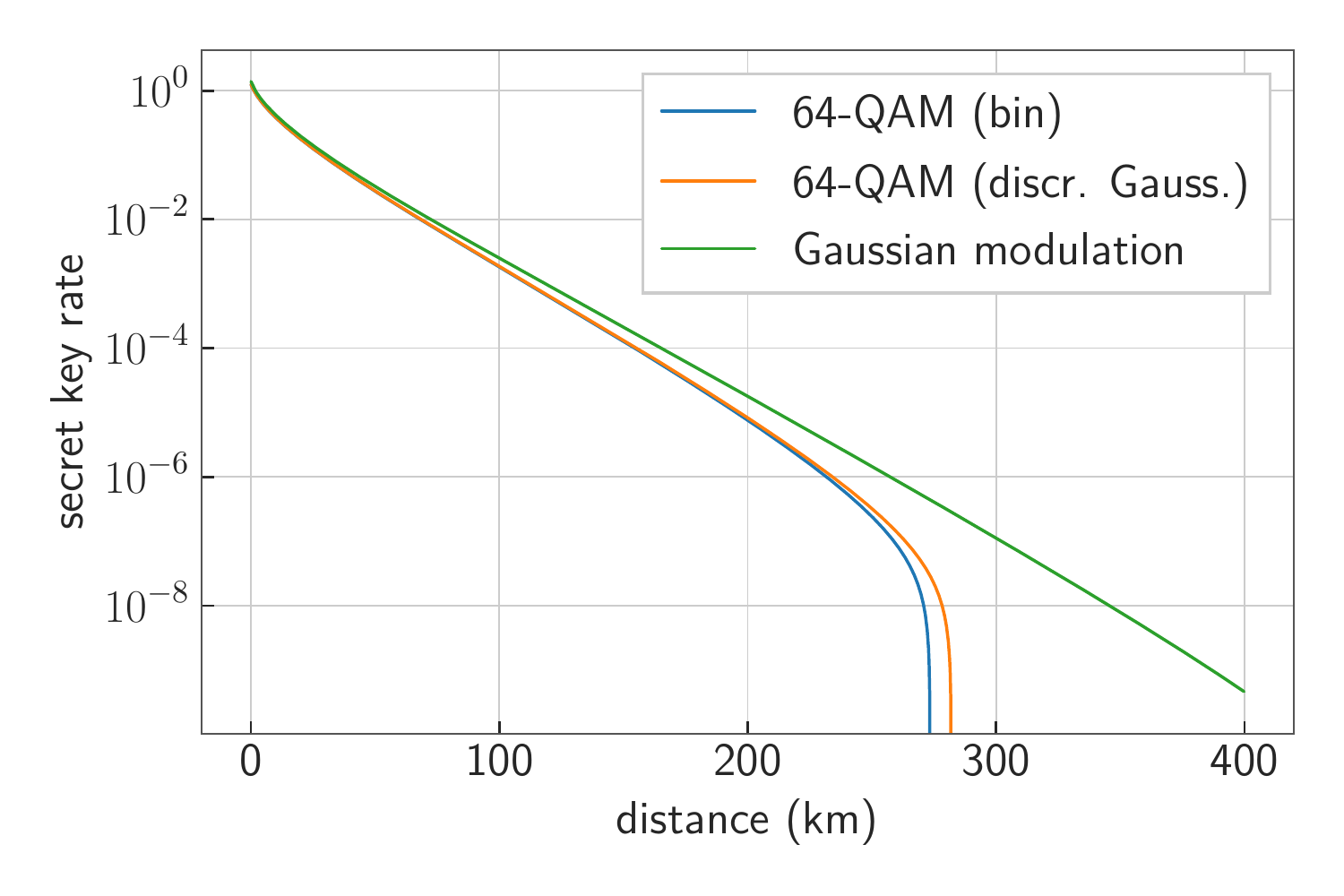}
\end{center}
\caption{Asymptotic secret key rate for the 16-QAM and 64-QAM, with two choices of distribution: binomial \textit{vs} discrete Gaussian. The fixed parameters are $V_A=5$, $\xi = 0.02$ and $\beta=0.95$. Left panel: 16-QAM ($\nu = 0.085$ for the discrete Gaussian distribution); right panel: 64-QAM ($\nu=0.07$ for the discrete Gaussian distribution). In both cases, the discrete Gaussian distribution outperforms the binomial distribution, but the difference is only significant for the 16-QAM.}
\label{fig:modulations}
\end{figure}

In Figure \ref{fig:modulations}, we compare the binomial and the discrete Gaussian distributions discussed in Section \ref{sec:constellation} in the case of the 16-QAM and the 64-QAM. Note that the two distributions coincide by construction for the 4-QAM (or QPSK modulation). It is clear that for a 64-QAM, both distributions yield essentially the same performance, which is close to that of a Gaussian modulation with the same variance. For the 16-QAM, however, the discrete Gaussian outperforms the binomial distribution, when the value of the parameter $\nu$ in Eqn.~\eqref{eqn:discrete-gauss} is optimized. This also suggests that there is still room for further improvement in the case of the 16-QAM (or maybe of the 32-QAM which we have not discussed here mostly because it would break the independence of the real and imaginary parts of Alice's variables, and therefore potentially complicate the reconciliation procedure), and that additional work might lead to the discovery of better modulation schemes. Let us still insist on the fact that here we assume that $\beta$ is equal to $0.95$, independently of the modulation scheme, but that reality is probably more complex. In other words, it is important to also consider the reconciliation procedure when optimizing the modulation scheme. 

\begin{figure}[!h]
\begin{center}
\includegraphics[,trim=0 0 0 2,clip,scale=0.6]{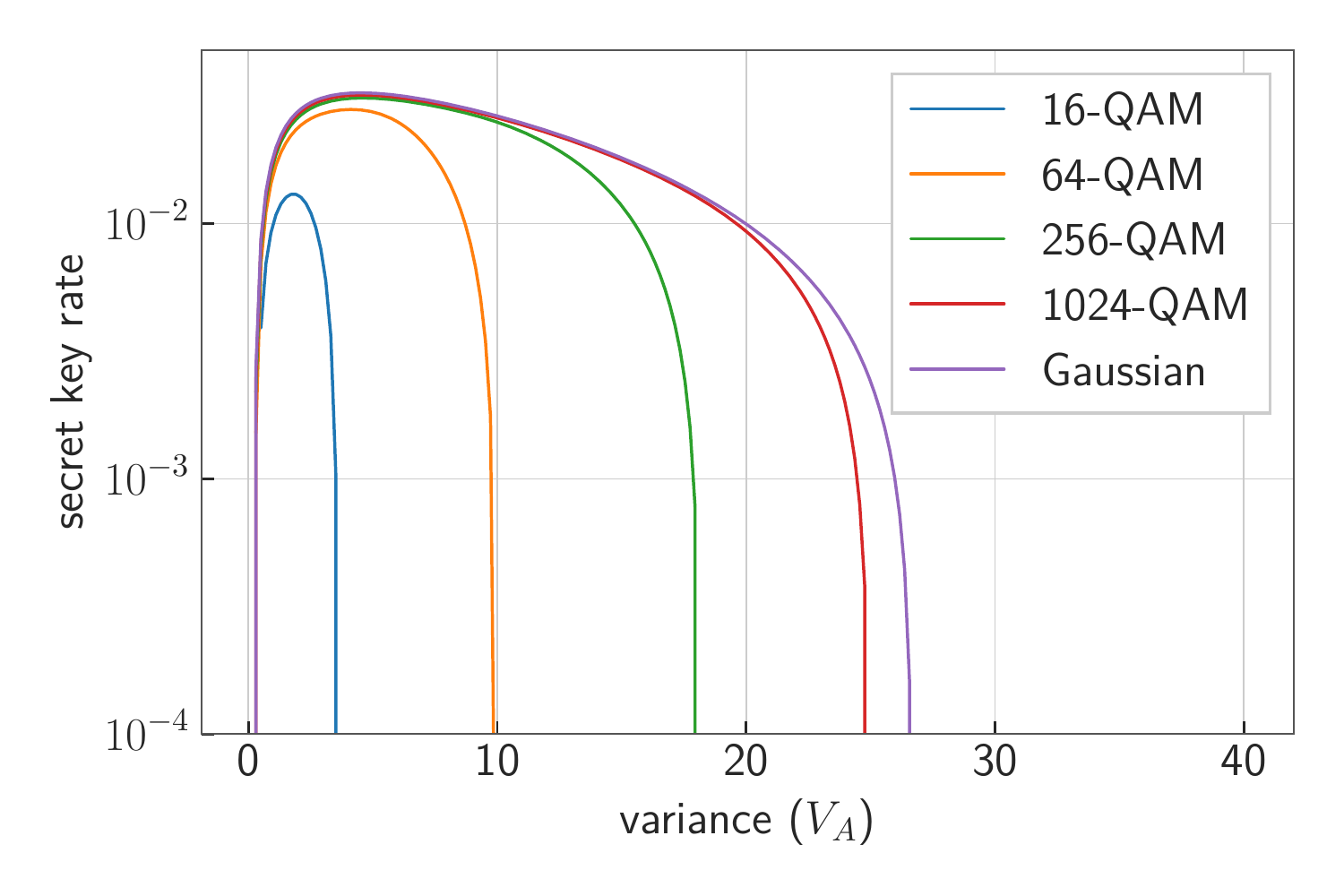}
\end{center}
\caption{Secret key rate at 50 km as a function of the modulation variance $V_A$, for various modulation schemes: from bottom to top: QAM of sizes $16, 64, 256, 1024$ (with the  binomial distribution of Eqn.~\eqref{eqn:bin} and \eqref{eqn:bin2}) and Gaussian modulation.  The other parameters are the excess noise $\xi = 0.02$ and the reconciliation efficiency $\beta = 0.95$. For this choice of distance and excess noise, our bound gives a vanishing secret key rate for the QPSK ($=$ 4-QAM).}
\label{fig:K-vs-VA}
\end{figure}

Figure \ref{fig:K-vs-VA} shows the performance of the various QAM sizes as a function of the modulation variance $V_A$. Here we only plot the results for the binomial distribution, since this avoids an extra optimization on $\nu$. The main observation is that increasing the size of the constellation brings the performance close to that of the Gaussian modulation for larger and larger values of $V_A$, allowing one to work at higher SNR, and thus simplifes the experimental implementation as well, possibly, as the reconciliation efficiency. At the same time, for a fixed reconciliation efficiency and a given distance (50 km here), we see that the optimal modulation variance is $V_A \approx 5$ and that the 64-QAM is already essentially indistinguishable from the Gaussian modulation.

\begin{figure}[!h]
\begin{center}
\includegraphics[trim=0 0 0 0,clip,scale=0.485]{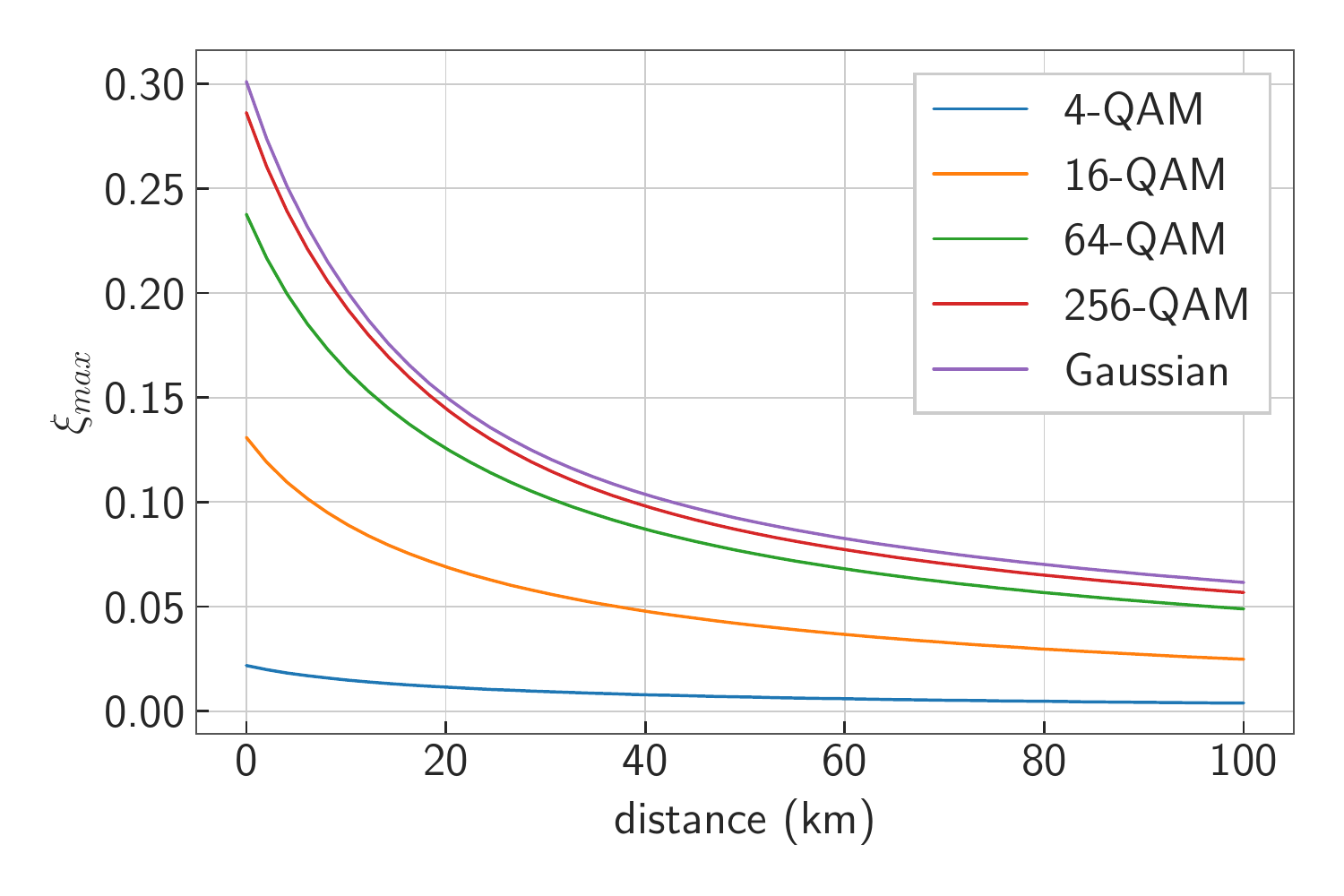}
\includegraphics[trim=0 0 0 0,clip,scale=0.485]{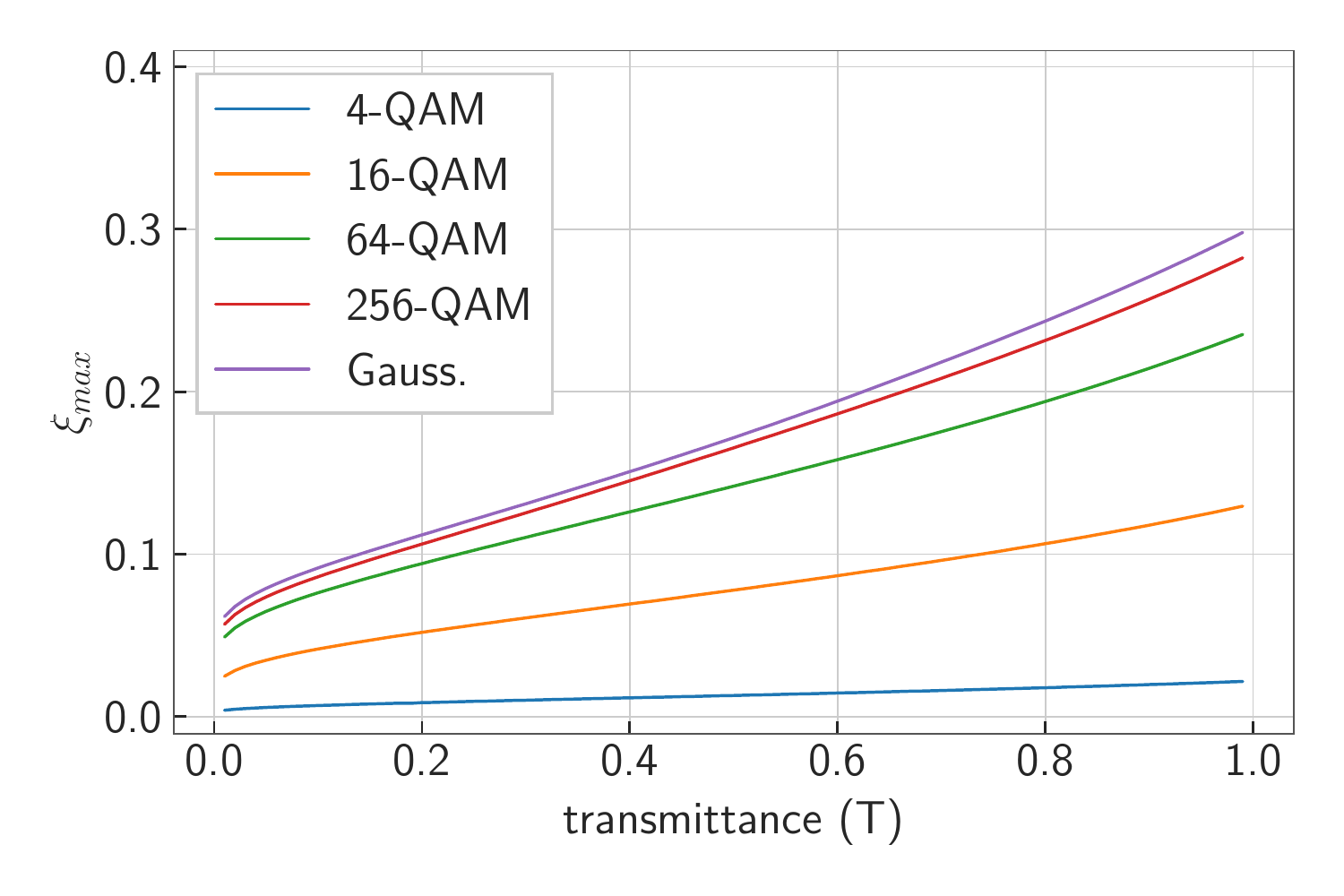}
\end{center}
\caption{Maximum value $\xi_{\max}$ of excess noise compatible with a positive key rate as a function of distance $d$ (left panel) or transmittance $T$ (right panel), for various QAM sizes (with binomial distribution). From bottom to top: 4-QAM to 256-QAM, and Gaussian modulation. The 1024-QAM (not displayed) is almost indistinguishable from the Gaussian modulation. Transmittance and distance are related through $T = 10^{-0.02d}$ with $d$ in km. Reconciliation efficiency is equal to $0.95$. The value of $V_A$ is optimized for each point.}
\label{fig:xi-max}
\end{figure}

Finally, we want to understand the performance of the various modulation schemes in terms of tolerable excess noise: if the transmittance of the channel is fixed to $T = 10^{-0.02d}$, what is the maximum value of the excess noise $\xi$ such that the secret key rate is positive? Figure \ref{fig:xi-max} shows the tolerable excess noise as a function of losses in the channel, when the modulation variance $V_A$ is optimized for each point. Again, we see that a 64-QAM already provides a performance close to the Gaussian modulation, and the 256-QAM is almost indistinguishable from the Gaussian modulation. The figures also confirm that our bound is quite bad for the QPSK modulation since the tolerable excess noise is at least an order of magnitude below what is achieved for larger QAM.

\paragraph{Acknowledgements.}
We thank Eleni Diamanti and Philippe Grangier for discussions on experimental considerations and modulation schemes, and Omar Fawzi for discussions on semidefinite programming. We also thank Fr\'ed\'eric Grosshans for his comments on a preliminary version of this manuscript, as well as an anonymous referee for suggesting simplified proofs in Sections \ref{sec:primal} and \ref{sec:thermal}.
AD and AL acknowledge funding from European Union’s Horizon’s Horizon 2020 Research and Innovation Programme under Grant Agreement No.~820466 (CiViQ). PB acknowledges funding from the European Research Council (ERC Grant Agreement No.~851716).

%\bibliographystyle{plainnat}
%\bibliography{CV}

\begin{thebibliography}{47}
\providecommand{\natexlab}[1]{#1}
\providecommand{\url}[1]{\texttt{#1}}
\expandafter\ifx\csname urlstyle\endcsname\relax
  \providecommand{\doi}[1]{doi: #1}\else
  \providecommand{\doi}{doi: \begingroup \urlstyle{rm}\Url}\fi

\bibitem[Bennett and Brassard(1984)]{BB84}
C.H. Bennett and G.~Brassard.
\newblock {Quantum cryptography: Public key distribution and coin tossing}.
\newblock In \emph{Proceedings of IEEE International Conference on Computers,
  Systems and Signal Processing}, volume 175, 1984.

\bibitem[Br{\'a}dler and Weedbrook(2018)]{BW18}
Kamil Br{\'a}dler and Christian Weedbrook.
\newblock Security proof of continuous-variable quantum key distribution using
  three coherent states.
\newblock \emph{Phys. Rev. A}, 97\penalty0 (2):\penalty0 022310, 2018.
\newblock \doi{10.1103/PhysRevA.97.022310}.

\bibitem[Cerf et~al.(2001)Cerf, Levy, and Van~Assche]{CLV01}
Nicolas~J Cerf, Marc Levy, and Gilles Van~Assche.
\newblock {Quantum distribution of Gaussian keys using squeezed states}.
\newblock \emph{Phys. Rev. A}, 63\penalty0 (5):\penalty0 052311, 2001.
\newblock \doi{10.1103/PhysRevA.63.052311}.

\bibitem[Christandl et~al.(2009)Christandl, K\"{o}nig, and Renner]{CKR09}
Matthias Christandl, Robert K\"{o}nig, and Renato Renner.
\newblock Postselection technique for quantum channels with applications to
  quantum cryptography.
\newblock \emph{Phys. Rev. Lett.}, 102\penalty0 (2):\penalty0 020504, 2009.
\newblock \doi{10.1103/PhysRevLett.102.020504}.

\bibitem[Denys et~al.(2021)Denys, Brown, and Leverrier]{DBL21}
Aur\'elie Denys, Peter Brown, and Anthony Leverrier.
\newblock Explicit asymptotic secret key rate of continuous-variable quantum
  key distribution with an arbitrary modulation of coherent states.
\newblock \emph{arXiv preprint arXiv:2011.09746v1}, 2021.

\bibitem[Devetak and Winter(2005)]{DW05}
I.~Devetak and A.~Winter.
\newblock {Distillation of secret key and entanglement from quantum states}.
\newblock In \emph{Proc. R. Soc. A}, volume 461, pages 207--235, 2005.
\newblock \doi{10.1098/rspa.2004.1372}.

\bibitem[Dupuis et~al.(2020)Dupuis, Fawzi, and Renner]{DFR20}
Frederic Dupuis, Omar Fawzi, and Renato Renner.
\newblock {Entropy accumulation}.
\newblock \emph{Communications in Mathematical Physics}, 379:\penalty0
  867–913, 2020.
\newblock \doi{10.1007/s00220-020-03839-5}.

\bibitem[Filip(2008)]{fil08}
Radim Filip.
\newblock Continuous-variable quantum key distribution with noisy coherent
  states.
\newblock \emph{Phys. Rev. A}, 77:\penalty0 022310, Feb 2008.
\newblock \doi{10.1103/PhysRevA.77.022310}.

\bibitem[Furrer et~al.(2012)Furrer, Franz, Berta, Leverrier, Scholz,
  Tomamichel, and Werner]{FFB12}
F.~Furrer, T.~Franz, M.~Berta, A.~Leverrier, V.~B. Scholz, M.~Tomamichel, and
  R.~F. Werner.
\newblock Continuous variable quantum key distribution: Finite-key analysis of
  composable security against coherent attacks.
\newblock \emph{Phys. Rev. Lett.}, 109:\penalty0 100502, 2012.
\newblock \doi{10.1103/PhysRevLett.109.100502}.

\bibitem[Garc\'{\i}a-Patr\'{o}n and Cerf(2006)]{GC06}
Ra\'{u}l Garc\'{\i}a-Patr\'{o}n and Nicolas~J. Cerf.
\newblock {Unconditional Optimality of Gaussian Attacks against
  Continuous-Variable Quantum Key Distribution}.
\newblock \emph{Phys. Rev. Lett.}, 97\penalty0 (19):\penalty0 190503, 2006.
\newblock \doi{10.1103/PhysRevLett.97.190503}.

\bibitem[Ghazisaeidi et~al.(2017)]{GJR17}
Amirhossein Ghazisaeidi et~al.
\newblock {Advanced C$+$L-Band Transoceanic Transmission Systems Based on
  Probabilistically Shaped PDM-64QAM}.
\newblock \emph{J. Lightwave Technol.}, 35\penalty0 (7):\penalty0 1291--1299,
  Apr 2017.
\newblock \doi{10.1109/JLT.2017.2657329}.

\bibitem[Ghorai et~al.(2019)Ghorai, Grangier, Diamanti, and Leverrier]{GGD19}
Shouvik Ghorai, Philippe Grangier, Eleni Diamanti, and Anthony Leverrier.
\newblock Asymptotic security of continuous-variable quantum key distribution
  with a discrete modulation.
\newblock \emph{Phys. Rev. X}, 9:\penalty0 021059, Jun 2019.
\newblock \doi{10.1103/PhysRevX.9.021059}.

\bibitem[Grosshans and Grangier(2002{\natexlab{a}})]{GG02b}
F.~Grosshans and P.~Grangier.
\newblock {Reverse reconciliation protocols for quantum cryptography with
  continuous variables}.
\newblock \emph{Arxiv preprint quant-ph/0204127}, 2002{\natexlab{a}}.

\bibitem[Grosshans et~al.(2003)Grosshans, Cerf, Wenger, Tualle-Brouri, and
  Grangier]{GCW03}
F.~Grosshans, N.J. Cerf, J.~Wenger, R.~Tualle-Brouri, and P.~Grangier.
\newblock {Virtual entanglement and reconciliation protocols for quantum
  cryptography with continuous variables}.
\newblock \emph{{Quantum Information and Computation}}, 3\penalty0 ({Sp. Iss.
  SI}):\penalty0 535--552, 2003.

\bibitem[Grosshans and Grangier(2002{\natexlab{b}})]{GG02}
Fr\'ed\'eric Grosshans and Philippe Grangier.
\newblock {Continuous Variable Quantum Cryptography Using Coherent States}.
\newblock \emph{Phys. Rev. Lett.}, 88\penalty0 (5):\penalty0 057902,
  2002{\natexlab{b}}.
\newblock \doi{10.1103/PhysRevLett.88.057902}.

\bibitem[Heid and L\"{u}tkenhaus(2007)]{HL07}
Matthias Heid and Norbert L\"{u}tkenhaus.
\newblock {Security of coherent-state quantum cryptography in the presence of
  Gaussian noise}.
\newblock \emph{Phys. Rev. A}, 76\penalty0 (2):\penalty0 022313, 2007.
\newblock \doi{10.1103/PhysRevA.76.022313}.

\bibitem[Hirano et~al.(2003)Hirano, Yamanaka, Ashikaga, Konishi, and
  Namiki]{HYA03}
Takuya Hirano, H~Yamanaka, M~Ashikaga, T~Konishi, and R~Namiki.
\newblock Quantum cryptography using pulsed homodyne detection.
\newblock \emph{Physical Review A}, 68\penalty0 (4):\penalty0 042331, 2003.
\newblock \doi{10.1103/PhysRevA.68.042331}.

\bibitem[Jardel et~al.(2018)Jardel, Eriksson, M{\'e}asson, Ghazisaeidi,
  Buchali, Idler, and Boutros]{JEM18}
Fanny Jardel, Tobias~A Eriksson, Cyril M{\'e}asson, Amirhossein Ghazisaeidi,
  Fred Buchali, Wilfried Idler, and Joseph~J Boutros.
\newblock Exploring and experimenting with shaping designs for next-generation
  optical communications.
\newblock \emph{Journal of Lightwave Technology}, 36\penalty0 (22):\penalty0
  5298--5308, 2018.
\newblock \doi{10.1109/JLT.2018.2871248}.

\bibitem[Jouguet et~al.(2011)Jouguet, Kunz-Jacques, and Leverrier]{JKL11}
Paul Jouguet, S\'ebastien Kunz-Jacques, and Anthony Leverrier.
\newblock {Long-distance continuous-variable quantum key distribution with a
  Gaussian modulation}.
\newblock \emph{Phys. Rev. A}, 84:\penalty0 062317, Dec 2011.
\newblock \doi{10.1103/PhysRevA.84.062317}.

\bibitem[Kaur et~al.(2021)Kaur, Guha, and Wilde]{KGW19}
Eneet Kaur, Saikat Guha, and Mark~M Wilde.
\newblock Asymptotic security of discrete-modulation protocols for
  continuous-variable quantum key distribution.
\newblock \emph{Physical Review A}, 103\penalty0 (1):\penalty0 012412, 2021.
\newblock \doi{10.1103/PhysRevA.103.012412}.

\bibitem[Lacerda et~al.(2016)Lacerda, Renes, and Scholz]{LRS16}
Felipe Lacerda, Joseph~M Renes, and Volkher~B Scholz.
\newblock {Coherent state constellations for Bosonic Gaussian channels}.
\newblock In \emph{Information Theory (ISIT), 2016 IEEE International Symposium
  on}, pages 2499--2503. IEEE, 2016.
\newblock \doi{10.1109/ISIT.2016.7541749}.

\bibitem[Leverrier(2015)]{lev15}
Anthony Leverrier.
\newblock Composable security proof for continuous-variable quantum key
  distribution with coherent states.
\newblock \emph{Phys. Rev. Lett.}, 114:\penalty0 070501, 2015.
\newblock \doi{10.1103/PhysRevLett.114.070501}.

\bibitem[Leverrier(2017)]{lev17}
Anthony Leverrier.
\newblock {Security of continuous-variable quantum key distribution via a
  Gaussian de Finetti reduction}.
\newblock \emph{Phys. Rev. Lett.}, 118:\penalty0 200501, May 2017.
\newblock \doi{10.1103/PhysRevLett.118.200501}.

\bibitem[Leverrier(2018)]{lev18}
Anthony Leverrier.
\newblock {SU(p, q) coherent states and a Gaussian de Finetti theorem}.
\newblock \emph{Journal of Mathematical Physics}, 59\penalty0 (4):\penalty0
  042202, 2018.
\newblock \doi{10.1063/1.5007334}.

\bibitem[Leverrier and Grangier(2009)]{LG09}
Anthony Leverrier and Philippe Grangier.
\newblock Unconditional security proof of long-distance continuous-variable
  quantum key distribution with discrete modulation.
\newblock \emph{Phys. Rev. Lett.}, 102:\penalty0 180504, May 2009.
\newblock \doi{10.1103/PhysRevLett.102.180504}.

\bibitem[Leverrier and Grangier(2011)]{LG11}
Anthony Leverrier and Philippe Grangier.
\newblock {Continuous-variable quantum-key-distribution protocols with a
  non-Gaussian modulation}.
\newblock \emph{Phys. Rev. A}, 83:\penalty0 042312, Apr 2011.
\newblock \doi{10.1103/PhysRevA.83.042312}.

\bibitem[Lin et~al.(2019)Lin, Upadhyaya, and L\"utkenhaus]{LUL19}
Jie Lin, Twesh Upadhyaya, and Norbert L\"utkenhaus.
\newblock Asymptotic security analysis of discrete-modulated
  continuous-variable quantum key distribution.
\newblock \emph{Phys. Rev. X}, 9:\penalty0 041064, Dec 2019.
\newblock \doi{10.1103/PhysRevX.9.041064}.

\bibitem[Lorenz et~al.(2004)Lorenz, Korolkova, and Leuchs]{LKL04}
S.~Lorenz, N.~Korolkova, and G.~Leuchs.
\newblock Continuous-variable quantum key distribution using polarization
  encoding and post selection.
\newblock \emph{Appl. Phys. B}, 79\penalty0 (3):\penalty0 273--277, 2004.
\newblock \doi{10.1007/s00340-004-1574-7}.

\bibitem[Mani et~al.(2021)Mani, Gehring, Grabenweger, \"Omer, Pacher, and
  Andersen]{MGP21}
Hossein Mani, Tobias Gehring, Philipp Grabenweger, Bernhard \"Omer, Christoph
  Pacher, and Ulrik~Lund Andersen.
\newblock Multiedge-type low-density parity-check codes for continuous-variable
  quantum key distribution.
\newblock \emph{Phys. Rev. A}, 103:\penalty0 062419, Jun 2021.
\newblock \doi{10.1103/PhysRevA.103.062419}.
\newblock URL \url{https://link.aps.org/doi/10.1103/PhysRevA.103.062419}.

\bibitem[Matsuura et~al.(2021)Matsuura, Maeda, Sasaki, and Koashi]{MMS21}
Takaya Matsuura, Kento Maeda, Toshihiko Sasaki, and Masato Koashi.
\newblock Finite-size security of continuous-variable quantum key distribution
  with digital signal processing.
\newblock \emph{Nature communications}, 12\penalty0 (1):\penalty0 1--13, 2021.
\newblock \doi{10.1038/s41467-020-19916-1}.

\bibitem[Milicevic et~al.(2018)Milicevic, Chen, Zhang, and Gulak]{MCZ18}
Mario Milicevic, Feng Chen, Lei~M Zhang, and P~Glenn Gulak.
\newblock {Quasi-cyclic multi-edge LDPC codes for long-distance quantum
  cryptography}.
\newblock \emph{NPJ Quantum Information}, 4:\penalty0 1--9, 2018.
\newblock \doi{10.1038/s41534-018-0070-6}.

\bibitem[Navascu\'{e}s et~al.(2006)Navascu\'{e}s, Grosshans, and
  Ac\'{\i}n]{NGA06}
Miguel Navascu\'{e}s, Fr\'{e}d\'{e}ric Grosshans, and Antonio Ac\'{\i}n.
\newblock {Optimality of Gaussian Attacks in Continuous-Variable Quantum
  Cryptography}.
\newblock \emph{Phys. Rev. Lett.}, 97\penalty0 (19):\penalty0 190502, 2006.
\newblock \doi{10.1103/PhysRevLett.97.190502}.

\bibitem[Papanastasiou and Pirandola(2021)]{PP21}
Panagiotis Papanastasiou and Stefano Pirandola.
\newblock {Continuous-variable quantum cryptography with discrete alphabets:
  Composable security under collective Gaussian attacks}.
\newblock \emph{Phys. Rev. Research}, 3:\penalty0 013047, Jan 2021.
\newblock \doi{10.1103/PhysRevResearch.3.013047}.

\bibitem[Pirandola et~al.(2020)Pirandola, Andersen, Banchi, Berta, Bunandar,
  Colbeck, Englund, Gehring, Lupo, Ottaviani, Pereira, Razavi, Shaari,
  Tomamichel, Usenko, Vallone, Villoresi, and Wallden]{PAB20}
S.~Pirandola, U.~L. Andersen, L.~Banchi, M.~Berta, D.~Bunandar, R.~Colbeck,
  D.~Englund, T.~Gehring, C.~Lupo, C.~Ottaviani, J.~L. Pereira, M.~Razavi,
  J.~Shamsul Shaari, M.~Tomamichel, V.~C. Usenko, G.~Vallone, P.~Villoresi, and
  P.~Wallden.
\newblock Advances in quantum cryptography.
\newblock \emph{Adv. Opt. Photon.}, 12\penalty0 (4):\penalty0 1012--1236, Dec
  2020.
\newblock \doi{10.1364/AOP.361502}.

\bibitem[Pirandola et~al.(2015)Pirandola, Ottaviani, Spedalieri, Weedbrook,
  Braunstein, Lloyd, Gehring, Jacobsen, and Andersen]{POS15}
Stefano Pirandola, Carlo Ottaviani, Gaetana Spedalieri, Christian Weedbrook,
  Samuel~L Braunstein, Seth Lloyd, Tobias Gehring, Christian~S Jacobsen, and
  Ulrik~L. Andersen.
\newblock High-rate measurement-device-independent quantum cryptography.
\newblock \emph{Nat. Photon.}, 9\penalty0 (6):\penalty0 397--402, 2015.
\newblock \doi{10.1038/nphoton.2015.83}.

\bibitem[Renner(2007)]{ren07}
R.~Renner.
\newblock Symmetry of large physical systems implies independence of
  subsystems.
\newblock \emph{Nat. Phys.}, 3\penalty0 (9):\penalty0 645--649, 2007.
\newblock \doi{10.1038/nphys684}.

\bibitem[Renner and Cirac(2009)]{RC09}
R.~Renner and J.~I. Cirac.
\newblock {de Finetti Representation Theorem for Infinite-Dimensional Quantum
  Systems and Applications to Quantum Cryptography}.
\newblock \emph{Phys. Rev. Lett.}, 102\penalty0 (11):\penalty0 110504, 2009.
\newblock \doi{10.1103/PhysRevLett.102.110504}.

\bibitem[Scarani et~al.(2009)Scarani, Bechmann-Pasquinucci, Cerf, Du{\v{s}}ek,
  L{\"u}tkenhaus, and Peev]{SBC08}
V.~Scarani, H.~Bechmann-Pasquinucci, N.~J. Cerf, M.~Du{\v{s}}ek,
  N.~L{\"u}tkenhaus, and M.~Peev.
\newblock The security of practical quantum key distribution.
\newblock \emph{Rev. Mod. Phys.}, 81\penalty0 (3):\penalty0 1301, 2009.
\newblock \doi{10.1103/RevModPhys.81.1301}.

\bibitem[Sych and Leuchs(2010)]{SL10}
Denis Sych and Gerd Leuchs.
\newblock Coherent state quantum key distribution with multi letter phase-shift
  keying.
\newblock \emph{New J. Phys.}, 12\penalty0 (5):\penalty0 053019, 2010.
\newblock \doi{10.1088/1367-2630/12/5/053019}.

\bibitem[Tomamichel and Renner(2011)]{TR11}
Marco Tomamichel and Renato Renner.
\newblock Uncertainty relation for smooth entropies.
\newblock \emph{Phys. Rev. Lett.}, 106:\penalty0 110506, Mar 2011.
\newblock \doi{10.1103/PhysRevLett.106.110506}.

\bibitem[Upadhyaya et~al.(2021)Upadhyaya, van Himbeeck, Lin, and
  L\"utkenhaus]{UHJ21}
Twesh Upadhyaya, Thomas van Himbeeck, Jie Lin, and Norbert L\"utkenhaus.
\newblock Dimension reduction in quantum key distribution for continuous- and
  discrete-variable protocols.
\newblock \emph{PRX Quantum}, 2:\penalty0 020325, 2021.
\newblock \doi{10.1103/PRXQuantum.2.020325}.

\bibitem[Usenko and Filip(2010)]{UF10}
Vladyslav~C. Usenko and Radim Filip.
\newblock Feasibility of continuous-variable quantum key distribution with
  noisy coherent states.
\newblock \emph{Phys. Rev. A}, 81:\penalty0 022318, Feb 2010.
\newblock \doi{10.1103/PhysRevA.81.022318}.

\bibitem[Weedbrook et~al.(2004)Weedbrook, Lance, Bowen, Symul, Ralph, and
  Lam]{WLB04}
Christian Weedbrook, Andrew~M. Lance, Warwick~P. Bowen, Thomas Symul,
  Timothy~C. Ralph, and Ping~Koy Lam.
\newblock Quantum cryptography without switching.
\newblock \emph{Phys. Rev. Lett.}, 93\penalty0 (17):\penalty0 170504, 2004.
\newblock \doi{10.1103/PhysRevLett.93.170504}.

\bibitem[Weedbrook et~al.(2010)Weedbrook, Pirandola, Lloyd, and Ralph]{WPL10}
Christian Weedbrook, Stefano Pirandola, Seth Lloyd, and Timothy~C. Ralph.
\newblock Quantum cryptography approaching the classical limit.
\newblock \emph{Phys. Rev. Lett.}, 105:\penalty0 110501, Sep 2010.
\newblock \doi{10.1103/PhysRevLett.105.110501}.

\bibitem[Weedbrook et~al.(2012)Weedbrook, Pirandola, Garc{\'i}a-Patr\'on, Cerf,
  Ralph, Shapiro, and Lloyd]{WPG12}
Christian Weedbrook, Stefano Pirandola, Ra\'ul Garc{\'i}a-Patr\'on, Nicolas~J.
  Cerf, Timothy~C. Ralph, Jeffrey~H. Shapiro, and Seth Lloyd.
\newblock Gaussian quantum information.
\newblock \emph{Rev. Mod. Phys.}, 84:\penalty0 621--669, 2012.
\newblock \doi{10.1103/RevModPhys.84.621}.

\bibitem[Wu and Verd{\'u}(2010)]{WV10}
Yihong Wu and Sergio Verd{\'u}.
\newblock {The impact of constellation cardinality on Gaussian channel
  capacity}.
\newblock In \emph{2010 48th Annual Allerton Conference on Communication,
  Control, and Computing (Allerton)}, pages 620--628, 2010.
\newblock \doi{10.1109/ALLERTON.2010.5706965}.

\bibitem[Zhao et~al.(2009)Zhao, Heid, Rigas, and L\"{u}tkenhaus]{ZHR09}
Yi-Bo Zhao, Matthias Heid, Johannes Rigas, and Norbert L\"{u}tkenhaus.
\newblock Asymptotic security of binary modulated continuous-variable quantum
  key distribution under collective attacks.
\newblock \emph{Phys. Rev. A}, 79:\penalty0 012307, 2009.
\newblock \doi{10.1103/PhysRevA.79.012307}.

\end{thebibliography}

\end{document}